\documentclass{iopart}

\expandafter\let\csname equation*\endcsname\relax
\expandafter\let\csname endequation*\endcsname\relax

\usepackage[acronym,shortcuts,sanitize=none]{glossaries}
\usepackage[colorlinks,linkcolor={blue},citecolor={blue},urlcolor={red}]{hyperref}
\usepackage{iopams}
\usepackage{subfig}
\usepackage[margin=10pt,font=small,format=hang]{caption}
\usepackage{multicol}
\usepackage{multirow}
\usepackage{array}
\usepackage{graphicx}
\usepackage[dvipsnames]{xcolor}
\usepackage{cleveref}
\usepackage{lineno}

\newcommand{\macro}[1]{\textcolor{black}{#1}}
\newcommand\CBCEVENTSIGMA{\macro{\ensuremath{5.1}}}
\newcommand\CBCEVENTIFAR{\macro{\ensuremath{203\,000}}}
\newcommand{\CHIRPDURATION}{\macro{\ensuremath{0.2}}} 
\newcommand\OBSEVENTFULLDATE{\macro{September 14, 2015 09:50:45 UTC}}
\newcommand\OBSEVENTAPPROXCOMBINEDSNR{\macro{\ensuremath{24}}}
\newcommand\CBCSECONDEVENTIFAR{\macro{2.3}} 
\newcommand{\MONESCOMPACT}{\macro{\ensuremath{36_{-4}^{+5}}}} 
\newcommand{\MTWOSCOMPACT}{\macro{\ensuremath{29_{-4}^{+4}}}} 
\newcommand{\DISTANCECOMPACT}{\macro{\ensuremath{410_{-180}^{+160}}}} 
\newcommand{\MFINALobsSIMPLE}{\macro{\ensuremath{70}}} 
\newcommand{\SPINFINALSIMPLE}{\macro{\ensuremath{0.7}}} 
\newcommand{\MONESCOMPACTSecondMonday}{\macro{\ensuremath{23_{-6}^{+18}}}} 
\newcommand{\MTWOSCOMPACTSecondMonday}{\macro{\ensuremath{13_{-5}^{+4}}}} 

\DeclareGraphicsExtensions{%
    .pdf,.PDF,%
    .png,.PNG,%
    .jpg,.mps,.jpeg,.jbig2,.jb2,.JPG,.JPEG,.JBIG2,.JB2}

\begin{document}

\pagenumbering{arabic}

\title[Noise characterization related to GW150914]{Characterization of transient noise in Advanced LIGO relevant to gravitational wave signal GW150914}

\author{%
B~P~Abbott$^{1}$,  
R~Abbott$^{1}$,  
T~D~Abbott$^{2}$,  
M~R~Abernathy$^{1}$,  
F~Acernese$^{3,4}$,
K~Ackley$^{5}$,  
M~Adamo$^{4,21}$,
C~Adams$^{6}$,  
T~Adams$^{7}$,
P~Addesso$^{3}$
% ...
\\
{(LIGO Scientific Collaboration and Virgo Collaboration)}%
}%
\medskip
\address {$^{\dag}$Deceased, May 2015. $^{\ddag}$Deceased, March 2015. $^{\sharp}$Deceased, May 2012.}%
\medskip
\address {$^{1}$LIGO, California Institute of Technology, Pasadena, CA 91125, USA }
\address {$^{2}$Louisiana State University, Baton Rouge, LA 70803, USA }
\address {$^{3}$Universit\`a di Salerno, Fisciano, I-84084 Salerno, Italy }
\address {$^{4}$INFN, Sezione di Napoli, Complesso Universitario di Monte S.Angelo, I-80126 Napoli, Italy }
\address {$^{5}$University of Florida, Gainesville, FL 32611, USA }
\address {$^{6}$LIGO Livingston Observatory, Livingston, LA 70754, USA }
\address {$^{7}$Laboratoire d'Annecy-le-Vieux de Physique des Particules (LAPP), Universit\'e Savoie Mont Blanc, CNRS/IN2P3, F-74941 Annecy-le-Vieux, France }
\address {$^{8}$Albert-Einstein-Institut, Max-Planck-Institut f\"ur Gravi\-ta\-tions\-physik, D-30167 Hannover, Germany }
\address {$^{9}$Nikhef, Science Park, 1098 XG Amsterdam, Netherlands }
\address {$^{10}$LIGO, Massachusetts Institute of Technology, Cambridge, MA 02139, USA }
\address {$^{11}$Instituto Nacional de Pesquisas Espaciais, 12227-010 S\~{a}o Jos\'{e} dos Campos, S\~{a}o Paulo, Brazil }
\address {$^{12}$INFN, Gran Sasso Science Institute, I-67100 L'Aquila, Italy }
\address {$^{13}$INFN, Sezione di Roma Tor Vergata, I-00133 Roma, Italy }
\address {$^{14}$Inter-University Centre for Astronomy and Astrophysics, Pune 411007, India }
\address {$^{15}$International Centre for Theoretical Sciences, Tata Institute of Fundamental Research, Bangalore 560012, India }
\address {$^{16}$University of Wisconsin-Milwaukee, Milwaukee, WI 53201, USA }
\address {$^{17}$Leibniz Universit\"at Hannover, D-30167 Hannover, Germany }
\address {$^{18}$Universit\`a di Pisa, I-56127 Pisa, Italy }
\address {$^{19}$INFN, Sezione di Pisa, I-56127 Pisa, Italy }
\address {$^{20}$Australian National University, Canberra, Australian Capital Territory 0200, Australia }
\address {$^{21}$The University of Mississippi, University, MS 38677, USA }
\address {$^{22}$California State University Fullerton, Fullerton, CA 92831, USA }
\address {$^{23}$LAL, Universit\'e Paris-Sud, CNRS/IN2P3, Universit\'e Paris-Saclay, 91400 Orsay, France }
\address {$^{24}$Chennai Mathematical Institute, Chennai 603103, India }
\address {$^{25}$Universit\`a di Roma Tor Vergata, I-00133 Roma, Italy }
\address {$^{26}$University of Southampton, Southampton SO17 1BJ, United Kingdom }
\address {$^{27}$Universit\"at Hamburg, D-22761 Hamburg, Germany }
\address {$^{28}$INFN, Sezione di Roma, I-00185 Roma, Italy }
\address {$^{29}$Albert-Einstein-Institut, Max-Planck-Institut f\"ur Gravitations\-physik, D-14476 Potsdam-Golm, Germany }
\address {$^{30}$APC, AstroParticule et Cosmologie, Universit\'e Paris Diderot, CNRS/IN2P3, CEA/Irfu, Observatoire de Paris, Sorbonne Paris Cit\'e, F-75205 Paris Cedex 13, France }
\address {$^{31}$Montana State University, Bozeman, MT 59717, USA }
\address {$^{32}$Universit\`a di Perugia, I-06123 Perugia, Italy }
\address {$^{33}$INFN, Sezione di Perugia, I-06123 Perugia, Italy }
\address {$^{34}$European Gravitational Observatory (EGO), I-56021 Cascina, Pisa, Italy }
\address {$^{35}$Syracuse University, Syracuse, NY 13244, USA }
\address {$^{36}$SUPA, University of Glasgow, Glasgow G12 8QQ, United Kingdom }
\address {$^{37}$LIGO Hanford Observatory, Richland, WA 99352, USA }
\address {$^{38}$Wigner RCP, RMKI, H-1121 Budapest, Konkoly Thege Mikl\'os \'ut 29-33, Hungary }
\address {$^{39}$Columbia University, New York, NY 10027, USA }
\address {$^{40}$Stanford University, Stanford, CA 94305, USA }
\address {$^{41}$Universit\`a di Padova, Dipartimento di Fisica e Astronomia, I-35131 Padova, Italy }
\address {$^{42}$INFN, Sezione di Padova, I-35131 Padova, Italy }
\address {$^{43}$CAMK-PAN, 00-716 Warsaw, Poland }
\address {$^{44}$University of Birmingham, Birmingham B15 2TT, United Kingdom }
\address {$^{45}$Universit\`a degli Studi di Genova, I-16146 Genova, Italy }
\address {$^{46}$INFN, Sezione di Genova, I-16146 Genova, Italy }
\address {$^{47}$RRCAT, Indore MP 452013, India }
\address {$^{48}$Faculty of Physics, Lomonosov Moscow State University, Moscow 119991, Russia }
\address {$^{49}$SUPA, University of the West of Scotland, Paisley PA1 2BE, United Kingdom }
\address {$^{50}$University of Western Australia, Crawley, Western Australia 6009, Australia }
\address {$^{51}$Department of Astrophysics/IMAPP, Radboud University Nijmegen, 6500 GL Nijmegen, Netherlands }
\address {$^{52}$Artemis, Universit\'e C\^ote d'Azur, CNRS, Observatoire C\^ote d'Azur, CS 34229, Nice cedex 4, France }
\address {$^{53}$MTA E\"otv\"os University, ``Lendulet'' Astrophysics Research Group, Budapest 1117, Hungary }
\address {$^{54}$Institut de Physique de Rennes, CNRS, Universit\'e de Rennes 1, F-35042 Rennes, France }
\address {$^{55}$Washington State University, Pullman, WA 99164, USA }
\address {$^{56}$Universit\`a degli Studi di Urbino ``Carlo Bo,'' I-61029 Urbino, Italy }
\address {$^{57}$INFN, Sezione di Firenze, I-50019 Sesto Fiorentino, Firenze, Italy }
\address {$^{58}$University of Oregon, Eugene, OR 97403, USA }
\address {$^{59}$Laboratoire Kastler Brossel, UPMC-Sorbonne Universit\'es, CNRS, ENS-PSL Research University, Coll\`ege de France, F-75005 Paris, France }
\address {$^{60}$Astronomical Observatory Warsaw University, 00-478 Warsaw, Poland }
\address {$^{61}$VU University Amsterdam, 1081 HV Amsterdam, Netherlands }
\address {$^{62}$University of Maryland, College Park, MD 20742, USA }
\address {$^{63}$Center for Relativistic Astrophysics and School of Physics, Georgia Institute of Technology, Atlanta, GA 30332, USA }
\address {$^{64}$Institut Lumi\`{e}re Mati\`{e}re, Universit\'{e} de Lyon, Universit\'{e} Claude Bernard Lyon 1, UMR CNRS 5306, 69622 Villeurbanne, France }
\address {$^{65}$Laboratoire des Mat\'eriaux Avanc\'es (LMA), IN2P3/CNRS, Universit\'e de Lyon, F-69622 Villeurbanne, Lyon, France }
\address {$^{66}$Universitat de les Illes Balears, IAC3---IEEC, E-07122 Palma de Mallorca, Spain }
\address {$^{67}$Universit\`a di Napoli ``Federico II,'' Complesso Universitario di Monte S.Angelo, I-80126 Napoli, Italy }
\address {$^{68}$NASA/Goddard Space Flight Center, Greenbelt, MD 20771, USA }
\address {$^{69}$Canadian Institute for Theoretical Astrophysics, University of Toronto, Toronto, Ontario M5S 3H8, Canada }
\address {$^{70}$Tsinghua University, Beijing 100084, China }
\address {$^{71}$Texas Tech University, Lubbock, TX 79409, USA }
\address {$^{72}$The Pennsylvania State University, University Park, PA 16802, USA }
\address {$^{73}$National Tsing Hua University, Hsinchu City, 30013 Taiwan, Republic of China }
\address {$^{74}$Charles Sturt University, Wagga Wagga, New South Wales 2678, Australia }
\address {$^{75}$University of Chicago, Chicago, IL 60637, USA }
\address {$^{76}$Caltech CaRT, Pasadena, CA 91125, USA }
\address {$^{77}$Korea Institute of Science and Technology Information, Daejeon 305-806, Korea }
\address {$^{78}$Carleton College, Northfield, MN 55057, USA }
\address {$^{79}$Universit\`a di Roma ``La Sapienza,'' I-00185 Roma, Italy }
\address {$^{80}$University of Brussels, Brussels 1050, Belgium }
\address {$^{81}$Sonoma State University, Rohnert Park, CA 94928, USA }
\address {$^{82}$Northwestern University, Evanston, IL 60208, USA }
\address {$^{83}$University of Minnesota, Minneapolis, MN 55455, USA }
\address {$^{84}$The University of Melbourne, Parkville, Victoria 3010, Australia }
\address {$^{85}$The University of Texas Rio Grande Valley, Brownsville, TX 78520, USA }
\address {$^{86}$The University of Sheffield, Sheffield S10 2TN, United Kingdom }
\address {$^{87}$University of Sannio at Benevento, I-82100 Benevento, Italy and INFN, Sezione di Napoli, I-80100 Napoli, Italy }
\address {$^{88}$Montclair State University, Montclair, NJ 07043, USA }
\address {$^{89}$Universit\`a di Trento, Dipartimento di Fisica, I-38123 Povo, Trento, Italy }
\address {$^{90}$INFN, Trento Institute for Fundamental Physics and Applications, I-38123 Povo, Trento, Italy }
\address {$^{91}$Cardiff University, Cardiff CF24 3AA, United Kingdom }
\address {$^{92}$National Astronomical Observatory of Japan, 2-21-1 Osawa, Mitaka, Tokyo 181-8588, Japan }
\address {$^{93}$School of Mathematics, University of Edinburgh, Edinburgh EH9 3FD, United Kingdom }
\address {$^{94}$Indian Institute of Technology, Gandhinagar Ahmedabad Gujarat 382424, India }
\address {$^{95}$Institute for Plasma Research, Bhat, Gandhinagar 382428, India }
\address {$^{96}$University of Szeged, D\'om t\'er 9, Szeged 6720, Hungary }
\address {$^{97}$Embry-Riddle Aeronautical University, Prescott, AZ 86301, USA }
\address {$^{98}$University of Michigan, Ann Arbor, MI 48109, USA }
\address {$^{99}$Tata Institute of Fundamental Research, Mumbai 400005, India }
\address {$^{100}$American University, Washington, D.C. 20016, USA }
\address {$^{101}$University of Massachusetts-Amherst, Amherst, MA 01003, USA }
\address {$^{102}$University of Adelaide, Adelaide, South Australia 5005, Australia }
\address {$^{103}$West Virginia University, Morgantown, WV 26506, USA }
\address {$^{104}$University of Bia{\l }ystok, 15-424 Bia{\l }ystok, Poland }
\address {$^{105}$SUPA, University of Strathclyde, Glasgow G1 1XQ, United Kingdom }
\address {$^{106}$IISER-TVM, CET Campus, Trivandrum Kerala 695016, India }
\address {$^{107}$Institute of Applied Physics, Nizhny Novgorod, 603950, Russia }
\address {$^{108}$Pusan National University, Busan 609-735, Korea }
\address {$^{109}$Hanyang University, Seoul 133-791, Korea }
\address {$^{110}$NCBJ, 05-400 \'Swierk-Otwock, Poland }
\address {$^{111}$IM-PAN, 00-956 Warsaw, Poland }
\address {$^{112}$Rochester Institute of Technology, Rochester, NY 14623, USA }
\address {$^{113}$Monash University, Victoria 3800, Australia }
\address {$^{114}$Seoul National University, Seoul 151-742, Korea }
\address {$^{115}$University of Alabama in Huntsville, Huntsville, AL 35899, USA }
\address {$^{116}$ESPCI, CNRS, F-75005 Paris, France }
\address {$^{117}$Universit\`a di Camerino, Dipartimento di Fisica, I-62032 Camerino, Italy }
\address {$^{118}$Southern University and A\&M College, Baton Rouge, LA 70813, USA }
\address {$^{119}$College of William and Mary, Williamsburg, VA 23187, USA }
\address {$^{120}$Instituto de F\'\i sica Te\'orica, University Estadual Paulista/ICTP South American Institute for Fundamental Research, S\~ao Paulo SP 01140-070, Brazil }
\address {$^{121}$University of Cambridge, Cambridge CB2 1TN, United Kingdom }
\address {$^{122}$IISER-Kolkata, Mohanpur, West Bengal 741252, India }
\address {$^{123}$Rutherford Appleton Laboratory, HSIC, Chilton, Didcot, Oxon OX11 0QX, United Kingdom }
\address {$^{124}$Whitman College, 345 Boyer Avenue, Walla Walla, WA 99362 USA }
\address {$^{125}$National Institute for Mathematical Sciences, Daejeon 305-390, Korea }
\address {$^{126}$Hobart and William Smith Colleges, Geneva, NY 14456, USA }
\address {$^{127}$Janusz Gil Institute of Astronomy, University of Zielona G\'ora, 65-265 Zielona G\'ora, Poland }
\address {$^{128}$Andrews University, Berrien Springs, MI 49104, USA }
\address {$^{129}$Universit\`a di Siena, I-53100 Siena, Italy }
\address {$^{130}$Trinity University, San Antonio, TX 78212, USA }
\address {$^{131}$University of Washington, Seattle, WA 98195, USA }
\address {$^{132}$Kenyon College, Gambier, OH 43022, USA }
\address {$^{133}$Abilene Christian University, Abilene, TX 79699, USA }
\address {$^{134}$Louisiana Tech University, Ruston, LA 71272, USA }



\begin{abstract}

On September 14, 2015, a gravitational wave signal from a coalescing black hole binary system was observed by the Advanced LIGO detectors. 
This paper describes the transient noise backgrounds used to determine the significance of the event (designated GW150914) and presents the results of investigations into potential correlated or uncorrelated sources of transient noise in the detectors around the time of the event.
The detectors were operating nominally at the time of GW150914. 
We have ruled out environmental influences and non-Gaussian instrument noise at either LIGO detector as the cause of the observed gravitational wave signal.

\end{abstract}


\maketitle


\section{Introduction}\label{sec:intro}

A gravitational wave signal, denoted GW150914, has been detected by the Advanced LIGO detectors \cite{Aasi:2016bh}. 
The recovered waveform indicated the source was a binary black hole system with component masses \MONESCOMPACT\ M$_\odot$ and \MTWOSCOMPACT\ M$_\odot$, which coalesced at a distance of \DISTANCECOMPACT\ Mpc away 
from Earth. 
The significance of the GW150914 event was measured to be greater than 
\CBCEVENTSIGMA\ $\sigma$, corresponding to a false-alarm rate of less than 1 event per 
\CBCEVENTIFAR\ years \cite{Aasi:2016bh}. 
The event, lasting \CHIRPDURATION\ seconds in Advanced LIGO's sensitive frequency range, was detected in independent searches for modeled compact binary coalescences (CBCs) and for unmodeled gravitational wave bursts \cite{Aasi:2016cbc, Aasi:2016burst}.

\begin{figure}
\centering
\includegraphics[width=\linewidth]{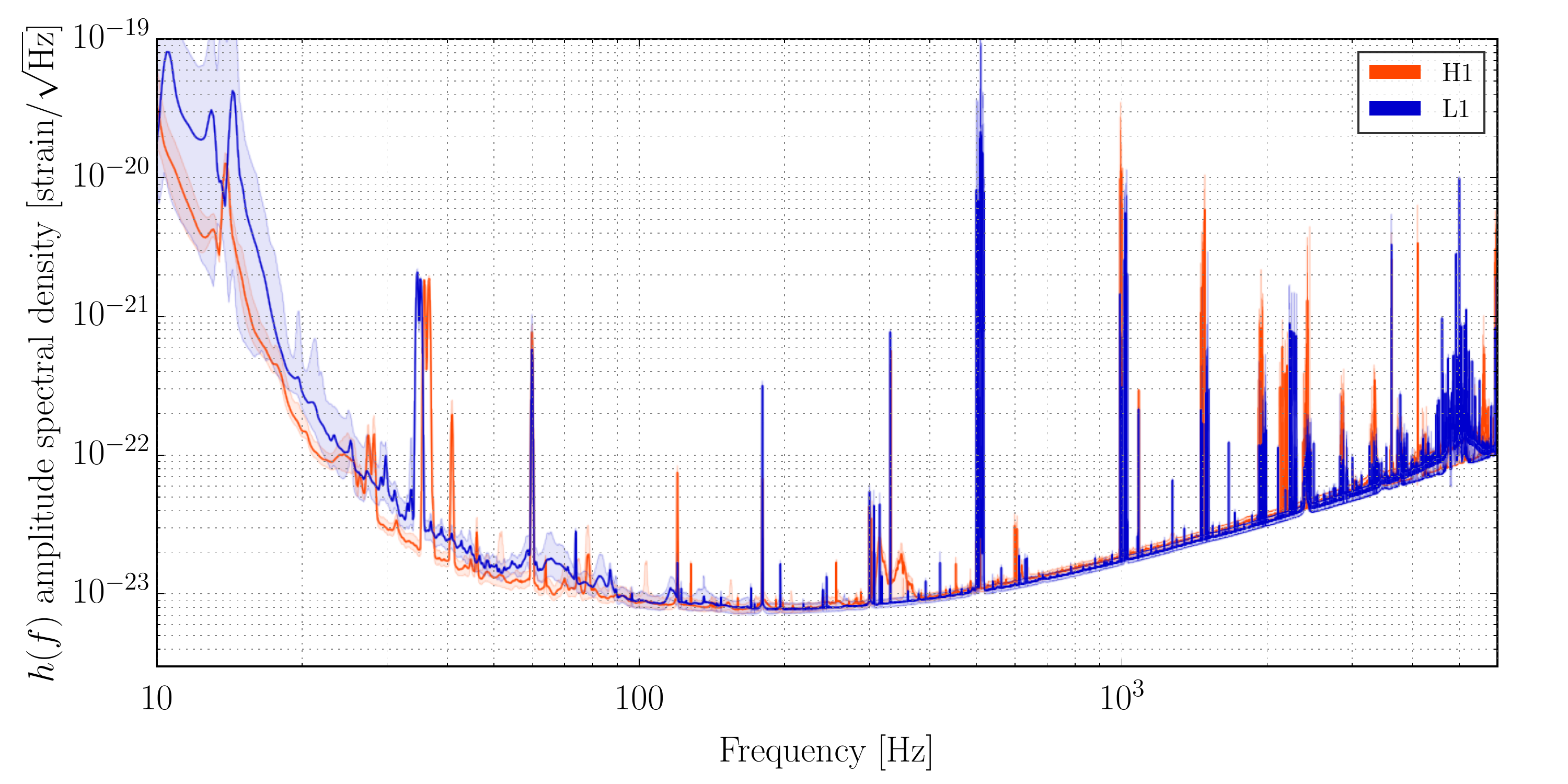}
\caption{The average measured strain-equivalent noise, or \textit{sensitivity}, of the Advanced LIGO detectors during the time analyzed to determine the significance of GW150914 (Sept 12 - Oct 20, 2015). LIGO-Hanford (H1) is shown in red, LIGO-Livingston (L1) in blue. The solid traces represent the median sensitivity and the shaded regions indicate the 5th and 95th percentile over the analysis period. The narrowband features in the spectra are due to known mechanical resonances, mains power harmonics, and injected signals used for calibration \cite{Aasi:2016in, Martynov:2016in, aLIGO}. 
}
\label{fig:aligo-asd}
\end{figure}

The US-based detectors, in Hanford, Washington (H1) and in Livingston, Louisiana (L1) jointly comprise the Laser Interferometer Gravitational-wave Observatory (LIGO). 
The detectors are designed to measure  spacetime strain induced by passing gravitational waves using a modified Michelson interferometer with 4~km length arms, as described in \cite{Aasi:2016in, Martynov:2016in, aLIGO}. 
The detectors were operating in their nominal configuration at the time of GW150914. 
The corresponding detector sensitivity is shown in Figure \ref{fig:aligo-asd}; both detectors achieved a best sensitivity of $\sim10^{-23}$ Hz$^{-1/2}$ between roughly 50 and 300 Hz. 
Peaks in the strain-equivalent noise amplitude spectral density are due largely to mechanical resonances, mains power harmonics, and injected signals used for calibration. 
Non-stationarity in the detector noise manifests as variations in the level and shape of these sensitivity curves over time.

Even in their nominal state, the detectors' data contain non-Gaussian noise transients introduced by behavior of the instruments or complex interactions between the instruments and their environment. 
For LIGO, the fundamental signature of a transient gravitational wave signal is a near-simultaneous signal with consistent waveforms in the two detectors.  
 The rate of coincident noise transients between the independent detector data sets is estimated by the astrophysical searches using time-shift techniques \cite{Aasi:2016cbc, Aasi:2016burst}. 
A common time-shift method is to shift the data of one detector relative to the other detector's data by a time interval significantly greater than 10~ms, the maximum difference in signal arrival time between detectors. 
Coincident triggers in time-shifted data yield a distribution of background triggers produced solely by the chance coincidence of transient noise. 
This time-shifting of the data is performed many times to obtain a representative estimate of the expected rate of background triggers, 
as detailed in \cite{Aasi:2016cbc, Usman:2015py}. The significance of a gravitational wave event is a measure of the probability that it is a false detection due to coincident noise. 
We study the characteristics of background triggers as well as correlations between the gravitational wave strain data and instrument or environment signals to guide further detector improvements and increase the sensitivity of the searches.

GW150914 occurred on \OBSEVENTFULLDATE, 
28 days into the eighth engineering run (ER8)\footnote{Engineering runs 1-7 served to test hardware and software infrastructure from the stability of instrument performance to the output of the astrophysical searches. ER8 was the final engineering run, intended to provide a gradual transition between a test of the mature instrument and search configurations and the continuous operation of an observing run.},
 3 days into stable data collection with an accurate calibration, and 4 days preceding the scheduled start of the first observing run (O1). 

After the event was identified as a highly significant candidate, the software and hardware configuration of each LIGO detector was held fixed until enough coincident data had been collected to set a sufficiently accurate upper bound on the false-alarm rate using the time-shift technique described above.  
It took roughly six weeks to collect the required $\sim$16 days of coincident data because low noise operation of the detectors is disrupted by noisy environmental conditions (such as storms, earthquakes, high ground motion, or anthropogenic noise sources). 
During this six week period we only performed non-invasive maintenance that was required for instrument stability.

The significance of GW150914 was calculated using data taken from September 12, 2015 00:00 through October 20, 2015 13:30 UTC.  
This data set was analyzed after removing time segments during which an identified instrumental or environmental noise source coupled to the gravitational wave strain signal.  At these times, any triggered output of the astrophysical searches would likely be due to noise. 
These \textit{data quality vetoes} were built on detector characterization efforts in earlier stages of testing and commissioning of the Advanced LIGO detectors, as reported in \cite{ER6DetChar}. 

This paper summarizes detector characterization techniques for identification of transient noise (Section \ref{sec:identifyingnoise}). We then present examples of transient noise couplings that can impact the detectors (Section \ref{sec:noisesources}) and discuss techniques used to mitigate the impact of known noise sources (Section \ref{sec:noisemitigation}). 
We show that the selected analysis period provides an accurate estimate of the significance of GW150914 reported in \cite{Aasi:2016bh} by discussing the stability of the search backgrounds, and presenting the impact of applied data quality vetoes relevant to GW150914 (Section \ref{sec:searchbackground}).
We also detail the specific checks performed to rule out an instrumental or
environmental noise-transient origin for GW150914, including potentially correlated
noise sources such as global magnetic noise that would not be captured by time-shift background estimation techniques (Section \ref{sec:gw150914}). 
Similar studies were also performed for the second most significant event in the CBC search over the analysis period, designated LVT151012\footnote{LIGO-Virgo Trigger (LVT) 151012 (October 12, 2015)}, observed~with a false alarm probability of $\sim$2\%~\cite{Aasi:2016bh, Aasi:2016cbc, Aasi:2016rates}.

%
\section{Identifying noise sources}\label{sec:identifyingnoise}

 In addition to the gravitational wave strain data, $h(t)$, each of the LIGO detectors also records over 200,000 \textit{auxiliary channels} that monitor instrument behavior and environmental conditions.
These channels witness a broad spectrum of potential coupling mechanisms, useful for diagnosing instrument faults and identifying noise correlations. 
Examples of instrument witness channels include measured angular drift of optics, light transmitted through a mirror as detected by a set of photodiodes, and actuation signals used to control optic position in order to maintain optical cavity resonance. 
In addition to candidate gravitational wave events, we study background triggers for correlation with trends or coincident transient noise in auxiliary channels on the broad scale of hours to days.  We also identify correlations on the order of the duration of transient astrophysical signals; a fraction of a millisecond to a few seconds. 
Systematic correlations are used to generate data quality vetoes used by the astrophysical searches to reduce the background, as described in \ref{sec:RF45}. 


An important set of auxiliary channels are the physical environment monitor (PEM) sensors, which monitor the local surroundings for potential disturbances that may affect the gravitational wave strain data, such as motion of the ground or optics tables, magnetic field variations, acoustic disturbances, or potentially, cosmic ray showers \cite{PEM}. 
A PEM sensor array is distributed throughout each detector site such that external environmental disturbances that could influence the detectors are witnessed with a significantly higher signal-to-noise ratio (SNR) in the PEM sensors than in $h(t)$. 
The PEM sensors are detailed in \ref{sec:PEM}.

\begin{figure}%
\centering
\includegraphics[width=\textwidth]{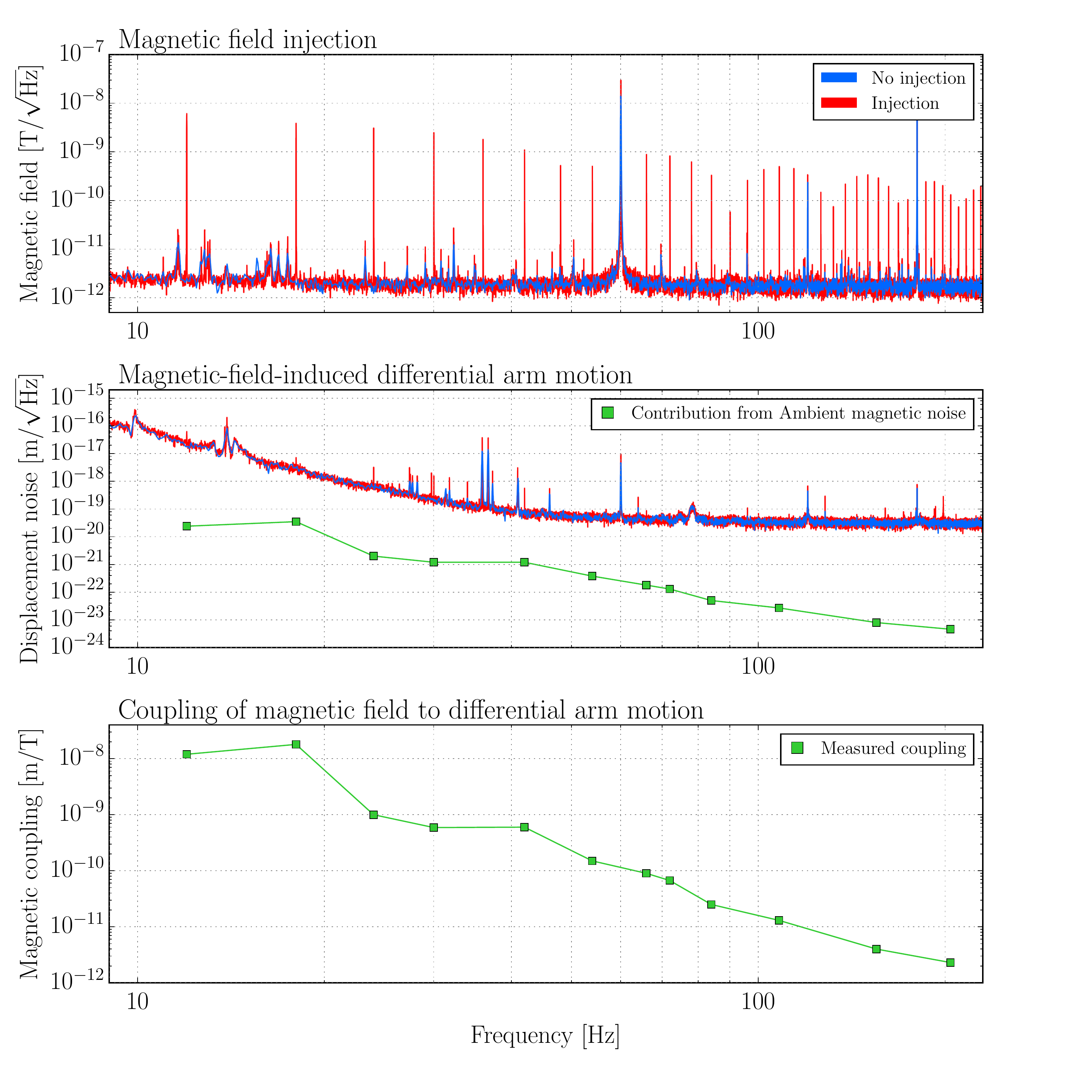}
  \caption{
  Noise coupling example: determining magnetic field coupling for a location at LIGO-Hanford. 
The top panel shows the output of a magnetometer installed in the corner station (see Figure \ref{fig:PEM})
during the injection of a series of single frequency oscillating magnetic fields at 6 Hz intervals (in red) and at a nominally quiet time (in blue). 
The middle panel shows $h(f)$ during this test (in red) and during the same nominally quiet time (in blue).
The heights of the induced peaks in $h(f)$ can be used to determine the magnetic coupling (in m/T) at those frequencies, as shown in the bottom panel.
The points in the bottom panel above 80 Hz were determined in a different test with a stronger magnetic field needed to produce discernible peaks in $h(f)$. 
The green points in the middle panel are an estimate of the contribution to $h(f)$ from the ambient magnetic noise during the nominally quiet time, calculated using the coupling function from the bottom panel. Injection tests also induced strong magnetic fields above 200~Hz. At higher frequencies, coupling was so low that the injected fields did not produce a response in $h(f)$, but were used to set upper limits on the coupling function.
This figure only shows data for one (typical) location, but similar injections were repeated at all locations where magnetic coupling might be of concern. 
  }
\label{fig:PEM_MAG_injections} 
\end{figure}


The relationship between environmental noise as witnessed by the PEM sensor array and the gravitational wave strain signal $h(t)$ is investigated using injection studies, where an intentional stimulus is introduced and the responses of both PEM sensors and the instrument are analyzed.
These injections ensure that the environmental sensors are more sensitive to environmental disturbances than the detector is, and also quantify the coupling between the environment and $h(t)$.
Figure \ref{fig:PEM_MAG_injections} illustrates a magnetic field injection test at the LIGO-Hanford detector that measured magnetic field coupling to $h(t)$ as well as the response of the local magnetometer to the injected field. 
The frequency-dependent coupling between the local magnetic field and $h(t)$ can be calculated from these measurements and used to accurately predict the response of $h(t)$ to the presence of a magnetic field, as witnessed by the local magnetometers. Figure \ref{fig:PEM_MAG_injections} shows an injection performed at one of the strongest coupling locations, in the building containing the beam splitter and most interferometer optics. Other magnetic field injection measurements identical to this test were also conducted for other locations throughout the detector site.
Similar injection studies were also conducted for radio, acoustic, and mechanical vibration sources. 

\section{Potential noise sources}\label{sec:noisesources}

Transient noise in $h(t)$ must occur within the frequency range targeted by the transient astrophysical searches to affect the background. This range is dictated by the equivalent strain noise of the detectors, as shown in \Cref{fig:aligo-asd} for the Hanford and Livingston detectors during the analysis period.

Motivated by this sensitivity curve, the transient astrophysical searches generally limit the search frequency range to above 30 Hz and below 2-3 kHz, or roughly the human-audible range. For example, a binary black hole signal like GW150914 is expected to have power measurable by the Advanced LIGO detectors between roughly 35 and 250 Hz and sources of short-duration noise with similar frequency content could impact the background estimation of such events. 

\subsection{Uncorrelated noise}

The following are examples of uncorrelated local noise features anticipated to be of particular interest or known to have a significant impact on the gravitational wave search backgrounds. 
The contribution of any uncorrelated noise sources is well estimated using time shifts.

\begin{itemize}
 
\item Some \textbf{anthropogenic noise} sources are likely to produce short duration transients in $h(t)$, such as human activity within one of the rooms that houses the vacuum chambers or infrequent strong ground motion or noise from other nearby locations. 
To reduce such vibrational or acoustic noise, detector staff do not enter the rooms containing the optical components of the detectors when the detectors are taking data. 
Any anthropogenic noise that could influence the detector is monitored by an array of accelerometers, seismometers, and microphones.
 
 \item \textbf{Earthquakes} can produce ground motion at the detectors with frequencies from approximately 0.03 to 0.1 Hz or higher if the epicenter is nearby \cite{PEM}. 
R-waves, the highest amplitude component of seismic waves from an earthquake \cite{eqwaves}, are the most likely to adversely impact data quality by rendering the detectors inoperable or inducing low frequency optic motion that up-converts to higher frequencies in $h(t)$ via mechanisms such as bilinear coupling of angular motion or light scattering \cite{Accadia:2010sc}.  
A network of seismometers installed at the LIGO detectors can easily identify earthquake disturbances. 

\item \textbf{Radio Frequency (RF) modulation} sidebands are used to sense and control a variety of optical cavities within the detector. 
Two modulations are applied to the input laser field at 9 and 45 MHz \cite{aLIGO}. 
Since the beginning of the analysis period, sporadic periods of a high rate of loud noise transients have been observed at LIGO-Hanford due to a fault in the 45 MHz electro-optic modulator driver system, which then couples to 
the gravitational wave channel between 10 and 2000 Hz, covering the entire frequency range analyzed by the CBC searches. 
Data associated with this electronic fault were vetoed and not analyzed. 
The engineering of this veto, as applied to the GW150914 analysis period, is detailed in \ref{sec:RF45}.

\item \textbf{Blip transients} are short noise transients that appear in the gravitational wave strain channel $h(t)$ as a symmetric `teardrop' shape in time-frequency space, typically between 30 and 250 Hz, with the majority of the power appearing at the lowest frequencies, as seen in Figure \ref{fig:blip_glitch}. 
They appear in both detectors independently with modest amplitude. 
The single detector burst identification algorithm Omicron, which identifies excess power transients using a generic sine-Gaussian time-frequency projection \cite{Robinet:2014om, Chatterji:2004q}, will resolve such noise transients with a signal-to-noise ratio of 10-100. 
No clear correlation to any auxiliary channel has yet been identified. As a result, there is currently no veto available to remove these noise transients from the astrophysical searches. 
Blip transients contribute to some of the most significant background triggers in both the unmodeled burst and modeled CBC searches. The noise transient shown in Figure \ref{fig:blip_glitch}\footnote{The spectrograms shown in Figures \ref{fig:blip_glitch}, \ref{fig:GW150914_omegascans}, and \ref{fig:G197392_omegascans} are generated using a sine-Gaussian basis \cite{Chattergi:2005om} instead of the sinusoidal basis of a traditional Fast-Fourier Transform.} is one example. 

\begin{figure}[!ht]%
\centering
  \includegraphics[width=0.495\textwidth]{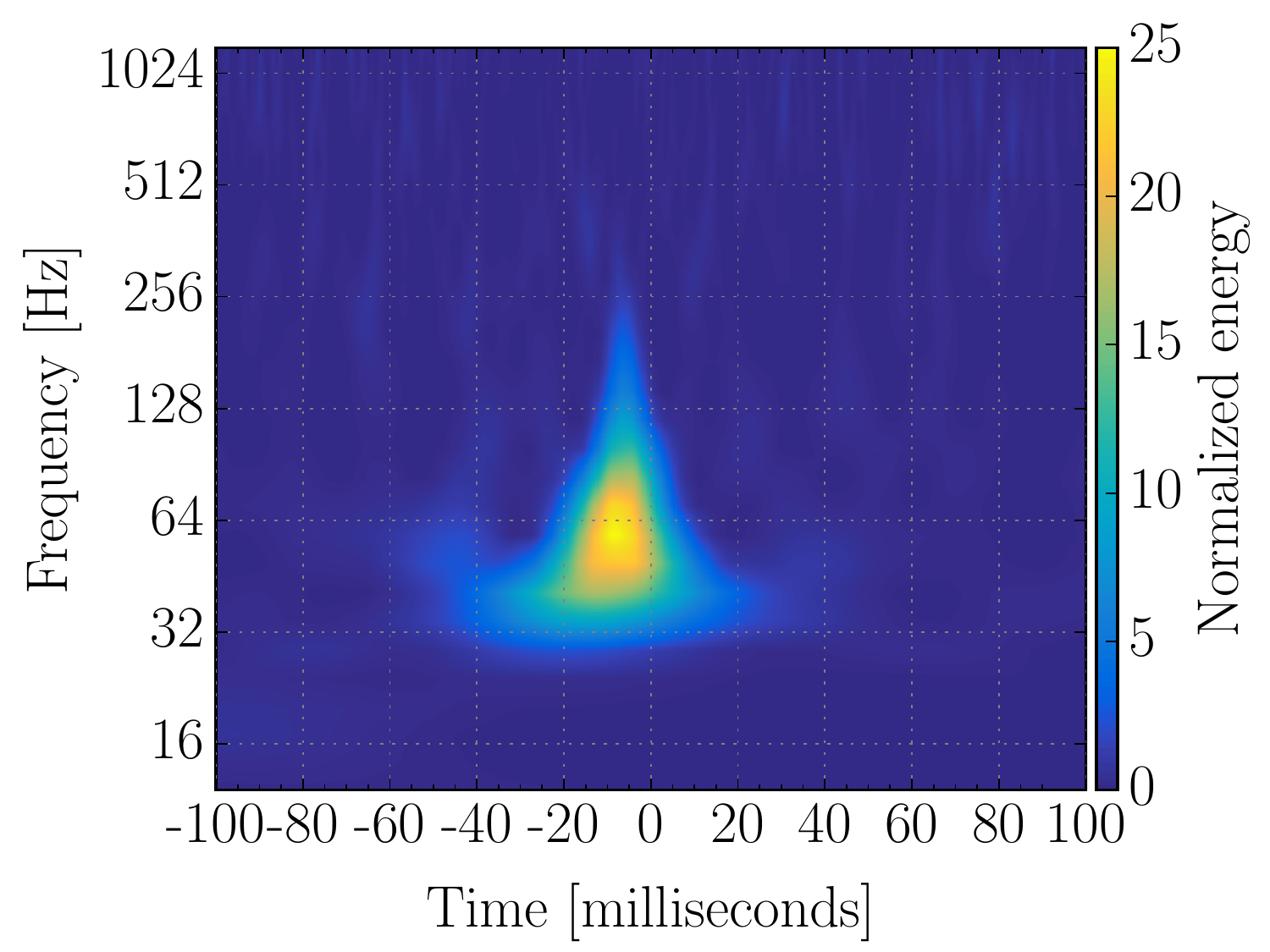}
  \caption{A normalized spectrogram of the LIGO-Livingston $h(t)$ channel at the time of a blip transient. The color scale indicates excess signal energy of data normalized by an estimated power spectral density.}
\label{fig:blip_glitch}
\end{figure}

\end{itemize}

The impact of noise sources on the astrophysical searches is discussed in Section~\ref{sec:DQvetoes}.

\subsection{Correlated noise}\label{sec:coincNoise}

Noise sources that may affect both detectors almost simultaneously could potentially imitate a gravitational wave event and would not be captured by time shifts in the search background estimation.

\textbf{Potential electromagnetic noise sources} include lightning, solar events and solar-wind driven noise, as well as radio frequency (RF) communication. If electromagnetic noise were strong enough to affect $h(t)$, it would be witnessed with high SNR by radio receivers and magnetometers. 

Lightning strikes occur tens of times per second globally. They can excite magnetic \textit{Schumann resonances}, a nearly harmonic series of peaks with a fundamental frequency near 8 Hz (governed by the light travel time around the earth) \cite{schumann:1952s1, schumann:1952s2}. However, the magnetic field amplitudes produced by Schumann resonances are of the order of a picoTesla; too small to produce strong signals in $h(t)$ (see Figure \ref{fig:PEM_MAG_injections}) \cite{Thrane:2013}. 

Nearby individual lightning strikes can induce transient noise in $h(t)$ via audio frequency magnetic fields generated by the lightning currents. However, even large strikes do not usually produce fields strong enough to be detected by the fluxgate magnetometers at both detectors simultaneously. 

Electromagnetic signals in the audio-frequency band are also produced by human and solar sources, including solar radio flares and currents of charged particles associated with the solar wind. The strongest solar or geomagnetic events during the analysis period were studied and no effect in $h(t)$ was observed at either detector. 

Electromagnetic fields that are outside the audio-frequency detection band are a potential concern because the LIGO detectors use RF modulation and demodulation for optical cavity control and because of the possibility of accidental demodulation with oscillators in the electronics systems. RF coupling measured during injection tests indicated that background RF fields were at least two orders of magnitude too small to influence the detector signal. 
The strongest coupling was found to be at the 9 and 45 MHz modulation frequencies used for control of optical cavities. These frequencies are monitored at both detectors with radio receivers that were at least two orders of magnitude more sensitive to fluctuations than the detector. 

\textbf{Cosmic ray showers} 
produce electromagnetic radiation and particle cascades when a highly energetic cosmic ray enters the Earth's atmosphere \cite{Braginsky;2006}. For even the most energetic showers, the 
cosmic ray flux drops effectively to zero within roughly 10 km of the axis of motion of the original collided particle \cite{Auger:2011as}, making coincident observation of a cosmic ray shower between the two detectors highly unlikely. 
As a precaution, a cosmic-ray detector is monitored at LIGO-Hanford; no coupling between cosmic ray particles and $h(t)$ has been observed.


\section{Mitigating noise sources}\label{sec:noisemitigation}

Ideally, when a noise source is identified, the instrument hardware or software is modified to reduce the coupling of the noise to $h(t)$ such that it no longer impacts astrophysical searches.
If mitigating the noise source is not viable, as in the case of data collected prior to an instrumental improvement, periods of time in which there are significant problems with the quality of the data are omitted, or \textit{vetoed}, from transient gravitational wave searches through a procedure similar to those utilized in previous LIGO analyses \cite{S6DetChar}. 

There are two different types of data quality products that can be applied as vetoes. Data quality \textit{flags} typically exclude periods of data on the order of seconds to hours when some reproducible criterion associated with known noise couplings is met \cite{S6DetChar, Christensen;2010, Amaldi, Macleod:2012sei}. 
For example, a data quality flag might be defined for periods when any of the photodiodes used to sense the laser field in the detector were overflowing their analog-to-digital converters. Data quality \textit{triggers} are short duration vetoes generated by algorithms that identify significant statistical correlations between a transient in $h(t)$ and transient noise in auxiliary channels \cite{Smith:2011, Isogai:2010, Essick:2013ov, Biswas:2013ap}. 

Data quality products are applied as vetoes in different categories that depend on the severity of the problem or the impact of individual data quality products on a search's background.  
Data quality flags used in \textit{category 1} collectively indicate times when data should not be analyzed due to a critical issue with a key detector component not operating in its nominal configuration. Since category-1-flagged times indicate major known problems with an instrument they are identically defined across all transient searches. 
Data quality flags used in \textit{category 2} collectively indicate times when a noise source with known physical coupling to $h(t)$ is active.  
Category 2 vetoes are typically applied after the initial processing of data for a specific search. This approach renders more data useable by the searches because they require unbroken strides of continuous data of up to 620 seconds for the coherent burst search and up to 2064 seconds for the CBC searches. There are three considerations for applying a data quality product as a category 2 veto to an astrophysical search: the physical noise coupling mechanism must be understood, the associated veto must have a demonstrated advantageous effect on the background of that search, and the veto must be \textit{safe}. 

The \textit{safety} of a veto is a measure of the likelihood that the veto criteria would accidentally remove a true gravitational wave signal. 
Veto safety is measured using hardware injection tests, where a signal is injected into $h(t)$ by inducing motion of the optics 
\cite{Smith:2011, Isogai:2010, Ajith;2014}.
If any auxiliary channels witness a corresponding response to a number of injected signals greater than expected by chance, these channels are considered \textit{unsafe} and are not used in the definition of any applied veto. 

The effectiveness of each data quality product in reducing the background is measured by the ratio of its \textit{efficiency}, or the fraction of background triggers it removes from a search, to its introduced \textit{deadtime}, or the fraction of time a particular flag will remove from the total duration of the set of analyzable data. Data quality flags used as category 2 vetoes have an efficiency-to-deadtime ratio for high SNR triggers significantly greater than 1, or the value expected for random behavior. An example is described in \ref{sec:RF45}.

A third veto category (category 3), applied in the same way as category 2, is generally reserved for data quality triggers, which are statistically generated, and data quality flags where the coupling mechanism is not understood. 
 
 During the GW150914 analysis period, data quality triggers 
 were applied as category 3 by burst searches. Times during hardware injection tests were also flagged and removed from the transient searches. 


Modeled CBC searches, which use matched filtering techniques \cite{Aasi:2016cbc}, apply additional mitigation methods to target loud noise transients with a duration on the order of a second or less that are particularly damaging. 
An accurate power spectral density (PSD) estimate is required to calculate the amount of signal power that matches a template waveform. 
Consequently, noise transients with a large amount of broadband power can corrupt the analyzed data up to the duration of the strain-equivalent noise PSD estimate, $\pm8$ seconds from the time of the noise transient. 
Additionally, a loud, short-duration noise transient can act as a delta function, which may imprint the impulse response of the matched filter on the output data, generating triggers. 
As a result, before analyzing the data the CBC searches apply a technique called \textit{gating} that smoothly rolls the input data stream off to zero for short-duration excursions identified as too loud to be consistent with an astrophysical signal \cite{Aasi:2016cbc}.


\section{Transient search backgrounds}\label{sec:searchbackground}

The data set used to calculate the significance of GW150914 is appropriate in both the stability of the search backgrounds over the analysis period and the judicious application of data quality vetoes. 

\subsection{Stability of the period analyzed for GW150914}

\begin{figure}
  \centering
  \includegraphics[width=\linewidth]{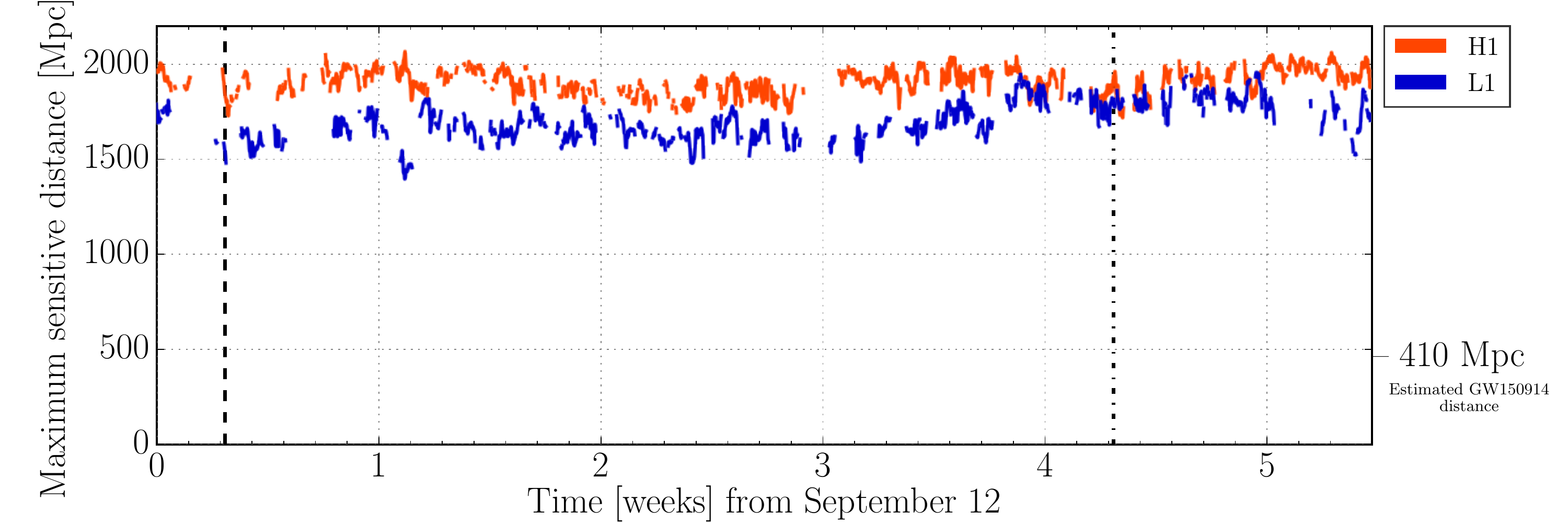}
  \caption{The maximum sensitivity of LIGO-Hanford (red) and LIGO-Livingston (blue) during the analyzed period (September 12 - %
           October 20 2015) to %
           a binary black hole system with the same observed spin and mass parameters as GW150914 for optimal sky location and source %
           orientation and detected with an SNR of 8. Each point was calculated using the PSD as %
           measured for each analysis segment (2048 seconds) of the CBC %
           search. The times of events GW150914 and LVT151012 are indicated %
           with vertical dashed and dot-dashed lines respectively. The LIGO-Livingston detector entered observation mode roughly 30 minutes prior to GW150914 after completing PEM injection tests in a stable, operational state. The LIGO-Hanford detector had been in observation mode for over an hour. }
  \label{fig:aligo-range}
\end{figure}

To illustrate the level of variability of detector performance over the several weeks of data collected for the analyzed time, \Cref{fig:aligo-range} shows the maximum sensitive distance of each of the detectors for 
the coalescence of a binary black hole system with the same spin and mass parameters as GW150914 in the detector frame (\MFINALobsSIMPLE\ M$_\odot$, \SPINFINALSIMPLE). This is calculated as the distance from Earth at which
the coalescence of a binary object pair produces an SNR of 8 in a single detector using matched filtering, assuming optimal sky location and source orientation. LIGO-Hanford had a mean maximum sensitive distance to GW150914-like signals of 1906 Mpc during the analysis period, and LIGO-Livingston had a mean of 1697 Mpc. 

\begin{figure}[!ht]%
\centering
  \includegraphics[width=\textwidth]{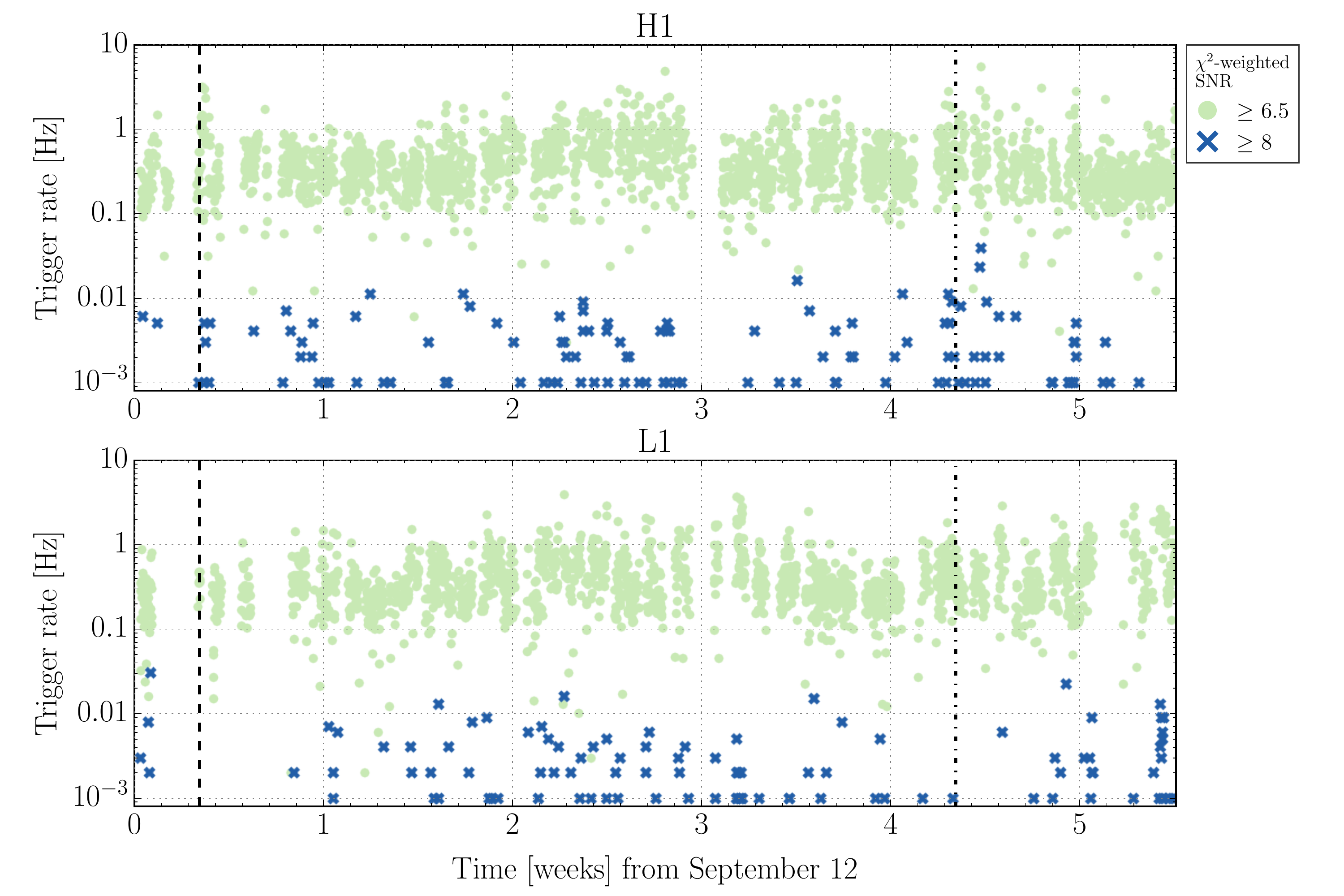}
  \caption{The rate of single interferometer background triggers in the CBC search for H1 %
           (above) and L1 (below), where color indicates a threshold on the detection statistic, $\chi^2$-weighted SNR. Each point represents the average rate over a 2048 second interval. 
           The times of GW150914 and LVT151012 are indicated with vertical dashed and dot-dashed lines respectively.
           }
  \label{fig:pycbc-trig-rates}%
\end{figure}

 LIGO-Hanford's maximum sensitive distance exhibited a 90\% range of $\sim$1800-2000 Mpc, and LIGO-Livingston's a 90\% range of $\sim$1500-1900,
 which was sufficiently stable to provide a reliable estimate of the CBC search background throughout the analysis period. 
 These small variations are due to a variety of fluctuations in the detectors and their environment, such as optic alignment variations or changing low frequency ground motion. 
 Figure \ref{fig:pycbc-trig-rates} shows the single-interferometer background trigger rate over time for the PyCBC search \cite{Usman:2015py} with two different thresholds on the detection statistic, $\chi^2$-weighted SNR\footnote{$\chi^2$-weighted SNR is the CBC detection statistic, where the SNR of a trigger is downweighted if there is excess power which does not match the template waveform.}  \cite{Aasi:2016cbc, Babek:2013cbc, Allen:Chisq}. 
Triggers with a $\chi^2$-weighted SNR $\geq$ 6.5 (shown in green) comprise the bulk of the distribution 
and indicate the overall trigger rate from the search: ${\sim}$1-10 Hz. Triggers with $\chi^2$-weighted SNR $\geq$ 8 (shown in blue) 
are fairly rare, typically showing up at a rate $<$ 
0.01 Hz during the analysis period.

\begin{figure}[!ht]%
\centering
  \includegraphics[width=\textwidth]{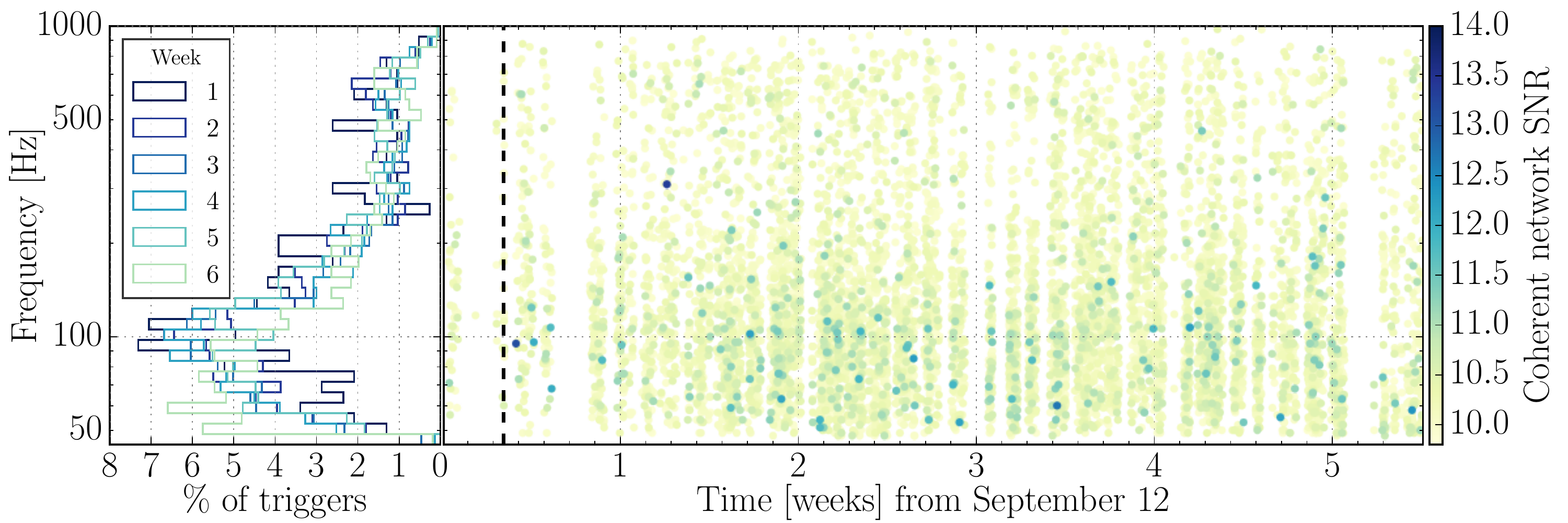}
  \caption{The behavior of cWB background triggers in frequency and coherent network SNR over the duration of the analysis period (right) and the frequency distribution of these triggers by week from September 12 to October 20, 2015 (left). For each time-shifted background trigger, the time for the Livingston detector is indicated. The time of GW150914, recovered with a coherent network SNR of 20, is indicated with a dashed vertical line in the right panel. (LVT151012 was not identified by cWB.) Overall, the background distribution is consistent throughout the analysis period.}
\label{fig:cwb_weekly_plots}
\end{figure}

The burst search background was also stable throughout the analysis containing GW150914. 
Figure \ref{fig:cwb_weekly_plots} shows the behavior of background triggers from the coherent all-sky burst search cWB (coherent WaveBurst)  \cite{Klimenko:2008, Klimenko:2015cWB2G} during the analysis period.  
In contrast to the single-interferometer CBC triggers shown in Figure \ref{fig:pycbc-trig-rates}, 
the coherent burst search requires coherent signal between multiple detectors to produce triggers, so the cWB background distribution is generated using time-shifted data. The main features of the background remain constant throughout the analyzed six weeks, particularly the domination of lower frequency triggers. 
Week 6 shows a small excess of triggers, $\sim$ 3$\%$ of total triggers, at lower than 60 Hz, which is below the majority of the power in event GW150914.

Variations in the environmental conditions and instrumental state throughout the analysis time, as captured in the range variation seen in Figure \ref{fig:aligo-range}, did not have a significant impact on the PyCBC or cWB background distributions. 

\subsection{The impact of data quality flags on the transient searches} \label{sec:DQvetoes}

\begin{figure}[!ht]%
\centering
  \subfloat[]{
      \includegraphics[width=.495\textwidth]{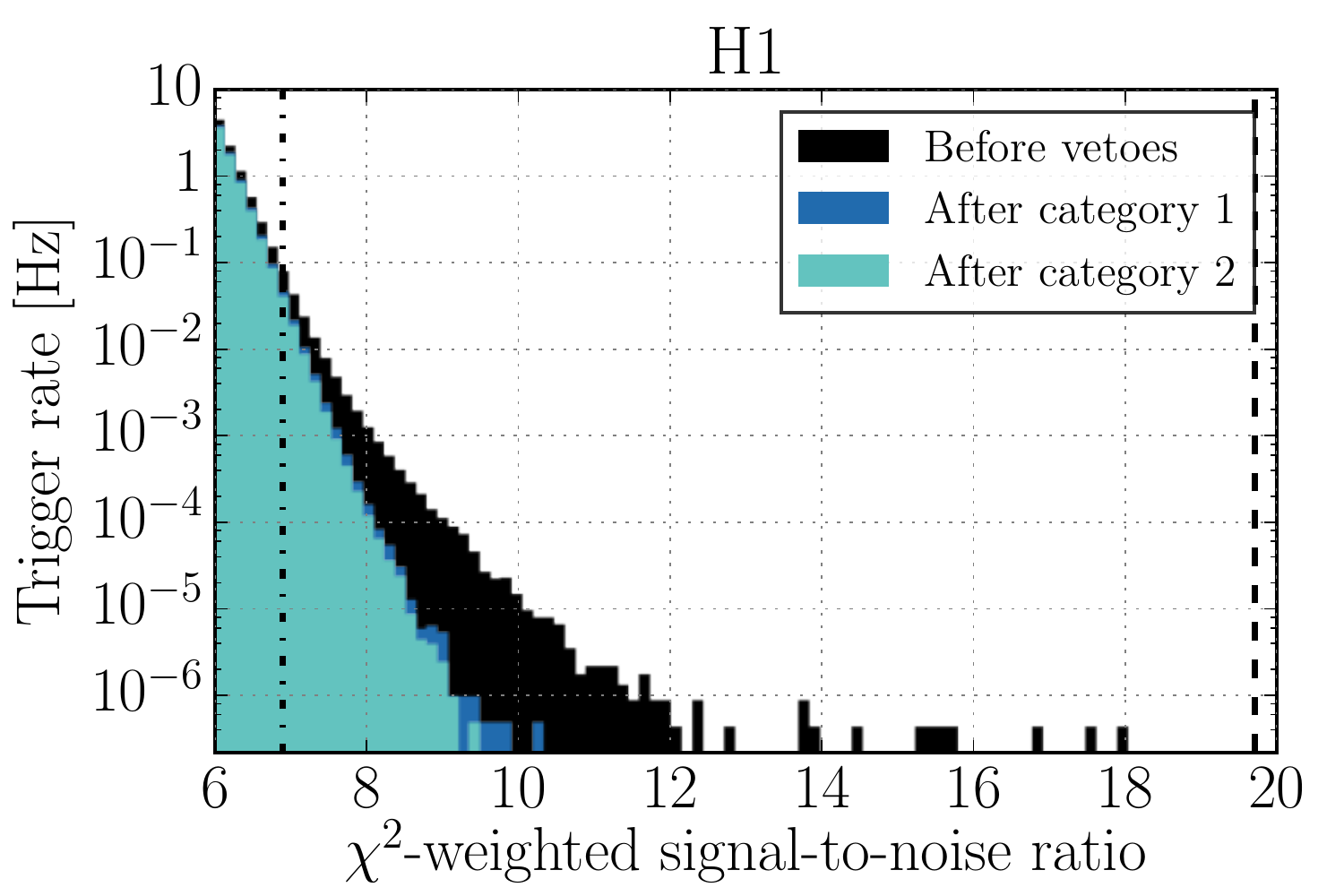}
      \label{subfig:h1-cbc-veto}
  }
  \subfloat[]{
      \includegraphics[width=.495\textwidth]{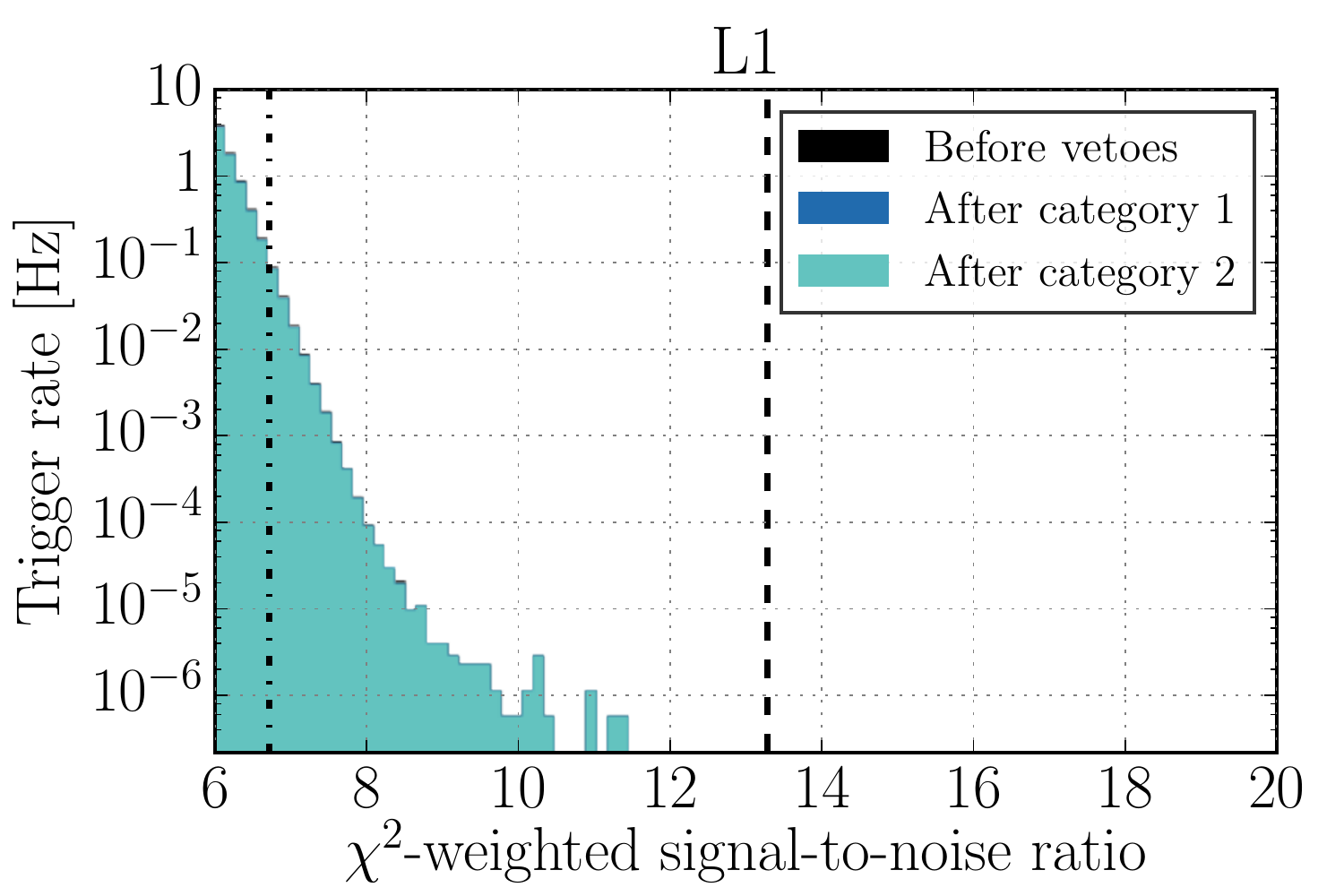}
      \label{subfig:l1-cbc-veto}
  }
  \caption{The impact of data-quality vetoes on the CBC background trigger %
           distribution for \protect\subref{subfig:h1-cbc-veto} LIGO-Hanford and %
           \protect\subref{subfig:l1-cbc-veto} LIGO-Livingston. The single-detector $\chi^2$-weighted SNR of GW150914 is indicated for each detector with a dashed line (19.7 for Hanford and 13.3 for Livingston), and for event LVT151012 with a dot-dashed line (6.9 for Hanford and 6.7 for Livingston).}
  \label{fig:cbc-newsnr-histograms}
\end{figure}

Data quality flags were generated independently for each detector in response to instrumental problems that demonstrated a well-defined, repeatable correlation with transient noise in $h(t)$.
Figure \ref{fig:cbc-newsnr-histograms} shows the CBC background trigger distributions from 
each detector with and without data quality products applied. 
The LIGO-Hanford background distribution was dramatically improved by the application of data quality vetoes, dominated by the effect of a single data quality flag. This flag was designed to indicate a fault in the phase modulation system used to create optical cavity control feedback signals, as discussed in \ref{sec:RF45}. LIGO-Livingston exhibits a longer tail of unvetoed background events which is largely composed of the blip noise transients discussed in Section \ref{sec:noisesources}.  
The total time removed from the CBC search by vetoes is summarized for each detector by veto category in Table \ref{table:CBCDQ}.

\begin{table}[!ht]
\parbox{.45\linewidth}{
\centering
\begin{tabular}{ccc}
 & Hanford & \\
 \hline
DQ veto & Total & \% of total   \\
category &deadtime (s) & coincident time  \\
\hline 
1 & 73446 &  4.62\% \\ 
2 & 5522 & 0.35\%  \\
\end{tabular}
}
\hfill
\parbox{.45\linewidth}{
\centering
\begin{tabular}{ccc}
 & Livingston & \\
 \hline
DQ veto & Total & \% of total   \\
category &deadtime (s) & coincident time  \\
\hline 
1 & 1066 & 0.07\%  \\ 
2 & 87 & 0.01\%  \\
\end{tabular}
}
\caption{The deadtime introduced by each data quality (DQ) veto category, as discussed in Section \ref{sec:noisemitigation}, for the CBC search during the analyzed period for LIGO-Hanford (left) and LIGO-Livingston (right).}
\label{table:CBCDQ}
\end{table}

For GW150914, the reported false-alarm probability was not significantly affected by these data quality vetoes.  GW150914 was the loudest recovered event during the analysis period -- significantly louder than every background event even without data quality products applied.  
 
For less significant triggers,  the application of data quality vetoes is more important \cite{Abbott;2016cbcdq}.  As an example, the false-alarm probability of the second most significant trigger (LVT151012) was 2\%.  Without the inclusion of data quality vetoes, the false-alarm probability would have been 14\%, increased by roughly a factor of 7. 

\begin{figure}[!ht]%
\centering
  \includegraphics[width=.5\textwidth]{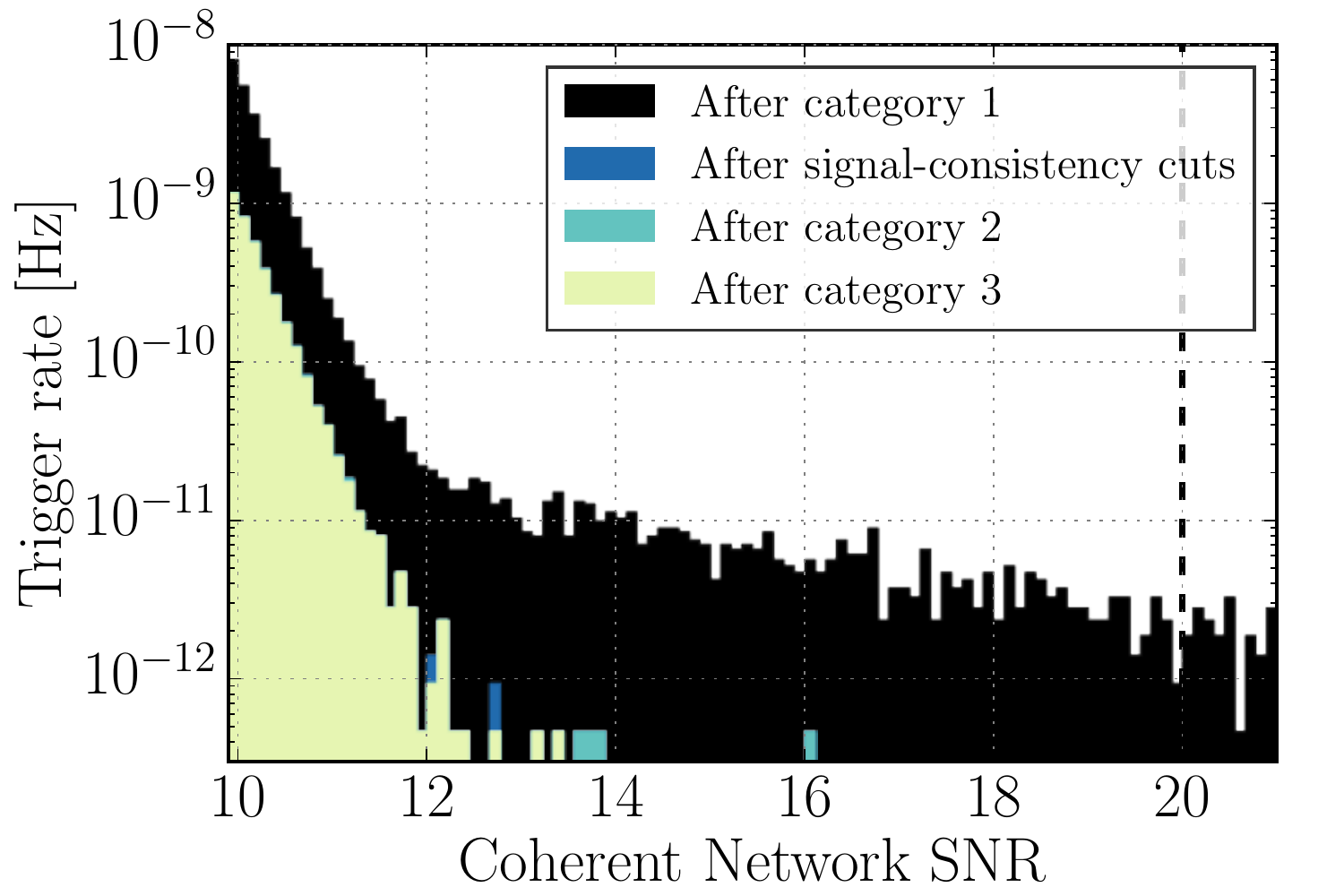}
  \caption{The impact of data-quality vetoes and signal consistency requirements on the background trigger %
           distribution from the cWB search for gravitational-wave bursts by coherent network SNR. The  multi-detector coherence required by cWB greatly reduces the rate of outlier events relative to the single-detector triggers shown in Figure \ref{fig:omicron-veto}. Note that the background rate is much lower than for single-interferometer triggers because it is normalized by the entire duration of the time-shifted analysis, not only the analysis period. The detected coherent network SNR of GW150914 is indicated with a dashed line. Note the background distributions shown here were selected to illustrate the effect of data quality vetoes and differ from those in Figure 4 of \cite{Aasi:2016bh}.
           }
  \label{fig:cwb-veto}
\end{figure}

\begin{table} [!ht]
\parbox{.45\linewidth}{
\centering
\begin{tabular}{ccc}
 & Hanford & \\
 \hline
DQ veto & Total & \% of total   \\
category &deadtime (s) & coincident time  \\
\hline 
1 & 73446 &  4.62\% \\ 
2 & 1900 & 0.12\%  \\
3 & 12815 & 0.81\%  \\
\end{tabular}
}
\hfill
\parbox{.45\linewidth}{
\centering
\begin{tabular}{ccc}
 & Livingston & \\
 \hline
DQ veto & Total & \% of total   \\
category &deadtime (s) & coincident time  \\
\hline 
1 & 1066 & 0.07\%  \\ 
2 & 736 & 0.05\%  \\
3 & 1319 & 0.08\%  \\
\end{tabular}
}
\caption{The deadtime introduced by each data quality (DQ) veto category for the coherent burst search during the analyzed period for LIGO-Hanford (left) and LIGO-Livingston (right).}
\label{table:cWBDQ}
\end{table}

\begin{figure}[!ht]
  \centering
  \subfloat[]{
    \includegraphics[width=.495\textwidth]{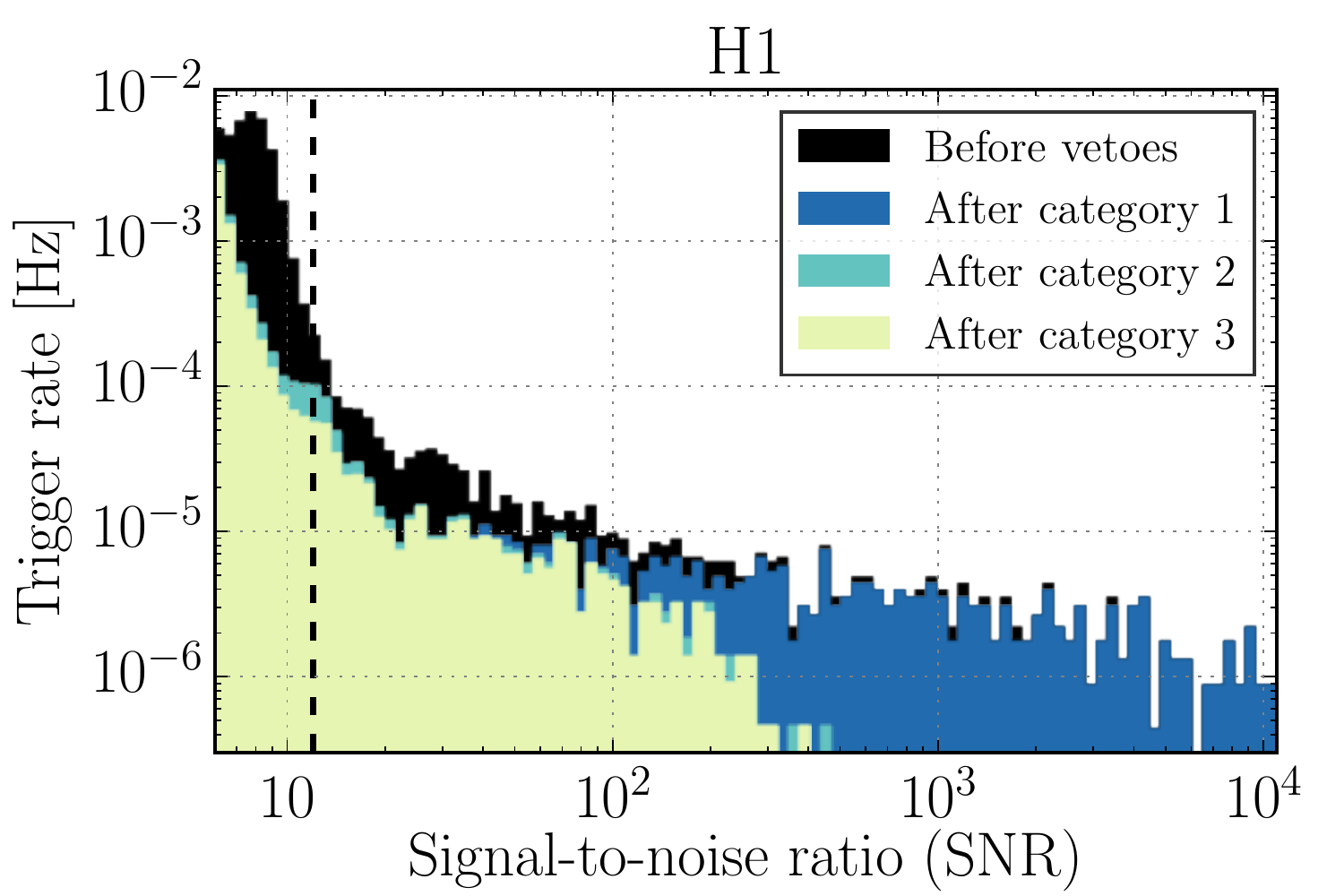}
    \label{fig:h1-omicron-veto}
  }
  \subfloat[]{
    \includegraphics[width=.495\textwidth]{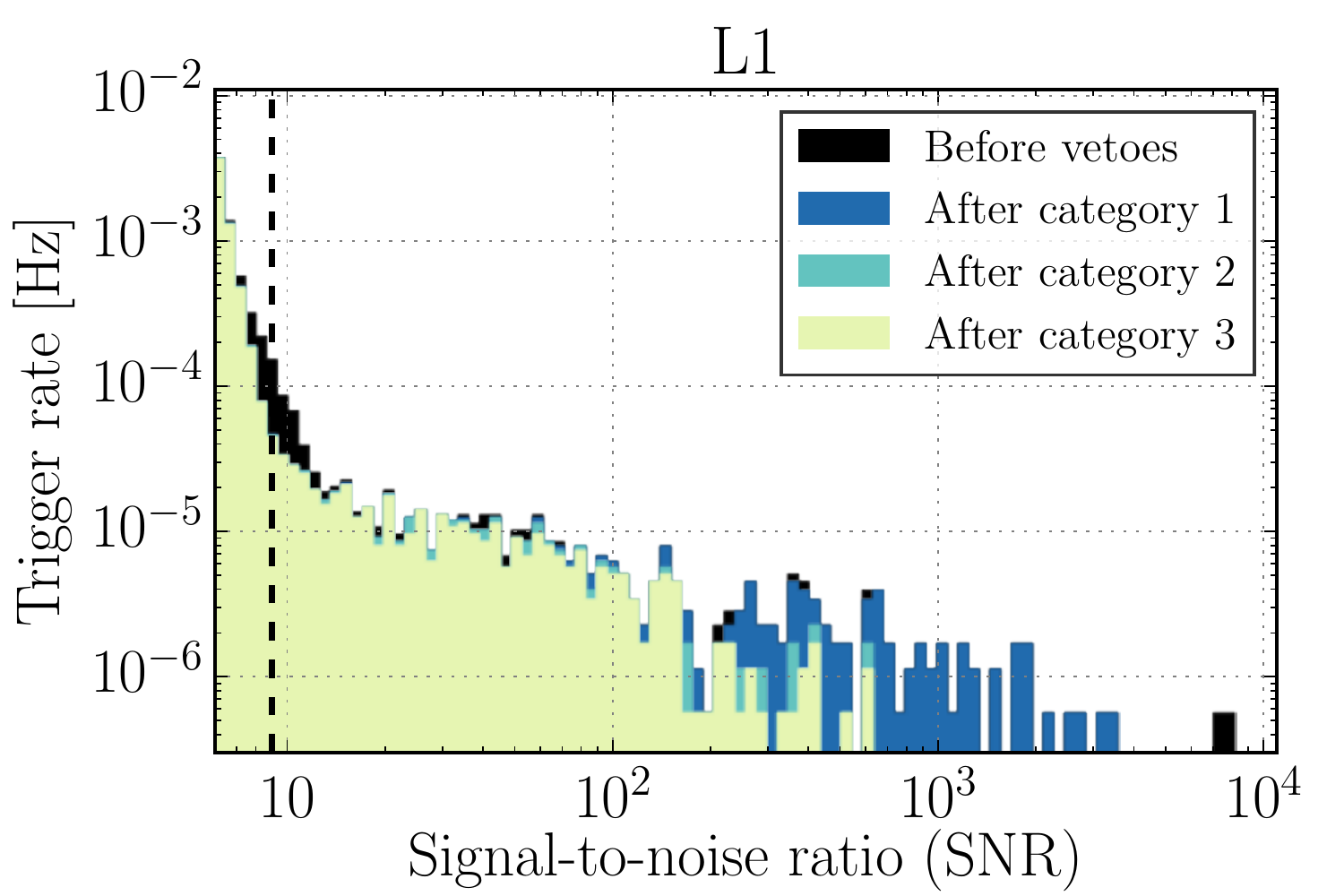}
    \label{fig:l1-omicron-veto}
  }
  \caption{The impact of data-quality vetoes on the single-detector burst %
           triggers detected by the Omicron burst algorithm for %
           \protect\subref{fig:h1-omicron-veto} LIGO-Hanford and %
           \protect\subref{fig:l1-omicron-veto} LIGO-Livingston. The SNR of GW150914 in each detector is indicated with a dashed line.}
  \label{fig:omicron-veto}
\end{figure}

Figure \ref{fig:cwb-veto} shows the impact of data-quality vetoes on the coherent burst search background, as well as the signal-consistency cut that requires resolved signals to have a time-frequency morphology consistent with expected astrophysical sources \cite{Aasi:2016burst}. 
The data quality flag with the highest efficiency-to-deadtime ratio for the coherent burst search background indicated large excursions in $h(t)$. This effective veto was defined using digital-to-analog overflows of the optic motion actuation signal used to stabilize the differential arm motion of the interferometer. 
This veto removed three of the loudest cWB background triggers during the analysis period. The remaining outliers with vetoes applied are blip-like noise transients of unknown instrumental origin. 

The total coincident time removed by each veto category from the burst search is summarized for each detector in Table \ref{table:cWBDQ}. Category 1 was defined identically between the burst and CBC searches, but there were some differences in the definition of category 2 largely due to differences in the observed impact of individual data quality products on the searches. 
For example, the CBC search used a data quality flag indicating periods of excess 10-30 Hz ground motion at LIGO-Hanford at category 2, but it was not applied to the burst search because it did not have a significant impact. The coherent burst search also applied a set of data quality triggers \cite{Smith:2011} at category 3, whereas the CBC search did not find this data quality product effective in reducing the background. 
A complete description of all data quality vetoes applied to the transient searches during the analysis period is reported in \cite{Nuttall:2016dq}.

Figure \ref{fig:omicron-veto} shows the effect of data quality vetoes on Omicron triggers from each detector. 
Since flags are tuned for specific problems at each detector, the impact on single-detector Omicron triggers is much more apparent than on the coherent burst search background in Figure \ref{fig:cwb-veto}, where the search requirement of a high degree of signal correlation between multiple detectors is effective in reducing the background. 

Figure \ref{fig:h1-omicron-veto} shows that the same category 1 data quality veto that dominated the reduction in the LIGO-Hanford CBC background distribution only impacted noise transients up to an SNR of roughly 100. 
The higher SNR Omicron triggers vetoed at category 2 from both detectors are mostly large excursions in $h(t)$ that are witnessed by overflows in the digital-to-analog conversion of the actuation signal controlling major optics, as mentioned for a data quality flag used effectively at category 2 for the coherent burst search. 
Blip noise transients are the main contributor to the unvetoed high SNR tail at both detectors along with 60-200 Hz nonstationarity that was persistent throughout the analysis period at LIGO-Livingston with an undetermined instrumental coupling.

%
\section{Transient noise around the time of GW150914}\label{sec:gw150914}

\begin{figure}[!ht]%
\centering
  \includegraphics[width=\textwidth]{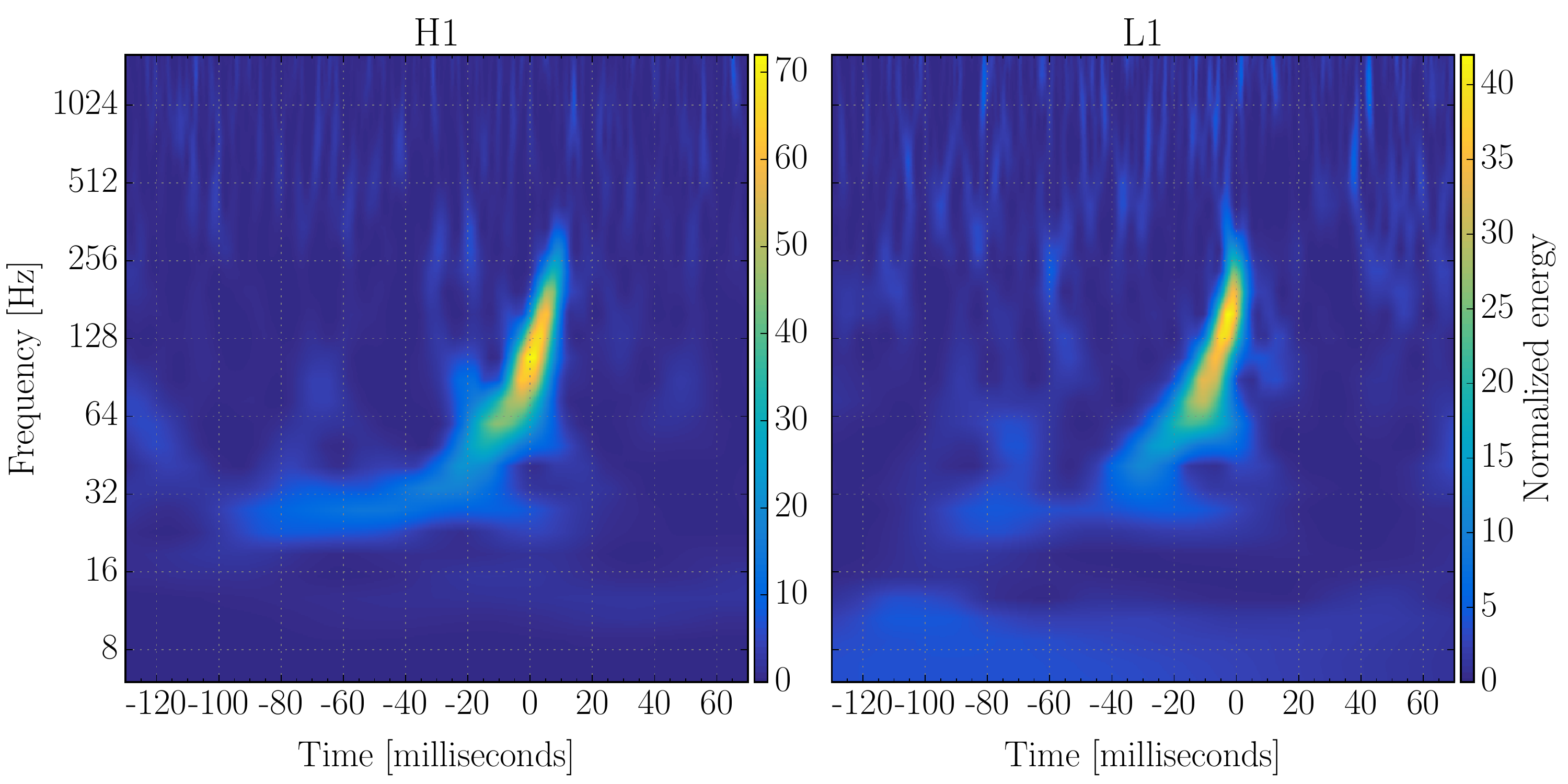}
  \caption{Normalized spectrograms
   of GW150914 in LIGO-Hanford (left) and LIGO-Livingston (right) $h(t)$ data with the same central GPS time.
  The data at both detectors exhibited typically low levels of noise around the time of the event; the signal, offset by $\sim$7~ms between detectors, was recovered by a matched-filter CBC search with a combined detector signal-to-noise ratio of \OBSEVENTAPPROXCOMBINEDSNR\ \cite{Aasi:2016bh, Aasi:2016cbc}, by the coherent burst search with a coherent network SNR of 20 \cite{Aasi:2016burst}, and by Omicron with a single-detector SNR of 12 in Hanford and 9 in Livingston.
   The time-frequency morphology of the event is distinct from the known noise sources discussed in Section \ref{sec:noisesources}. }
\label{fig:GW150914_omegascans} 
\end{figure}

The GW150914 event produced a strong gravitational wave signal in the Advanced LIGO detectors that shows the expected form of a binary black hole coalescence, as shown in Figure \ref{fig:GW150914_omegascans} \cite{Aasi:2016bh, Aasi:2016gr}. Immediately around the event the data are clean and stationary. 

Even though the routine data quality checks did not indicate any problems with the data, in-depth checks of potential noise sources were performed around the time of GW150914.
Potential noise couplings were considered from sources internal to the detector and local to each site, as well as common, coincident sources external to the detectors. 
All checks returned negative results for any pollution or interference large enough to have caused GW150914. 
Activities of personnel at the detectors, both locally and via remote internet connections, were confirmed to have no potential to induce transient noise in $h(t)$. 
Because GW150914 occurred during the early morning hours at both detectors, the only people on-site were the control room operators.
Signs of any anomalous activity nearby and the state of signal hardware injections were also investigated. 
These checks came back conclusively negative \cite{Aasi:2016site}.  
No data quality vetoes were active within an hour of the event.  Rigorous checks of the data calibration 
were also performed \cite{Aasi:2016cal}.

The results of a key subset of checks intended to demonstrate nominal detector performance, quiet environment behavior, and clean data quality around the event are reported here.  

For example, the U.S.\ Geological Survey (USGS) \cite{USGS} reported 
two magnitude 2.1 earthquakes within 20 minutes of GW150914; one with an epicenter off the coast of Alaska and another 70 miles south-west of Seattle.
The earthquakes produced minimal vertical ground motion at 0.03-0.1 Hz at the time of arrival; roughly 10~nm/s as measured by local seismometers at both detectors, which is an order of magnitude too small to produce an impact on the detector data. 

\subsection{Checks for potentially coincident noise sources}

The primary means of detecting the rare electromagnetic events that could conceivably produce coincident noise between the detectors are the array of magnetometers and radio receivers at each detector. 
These and all other PEM sensors were checked for 1 second around the time of GW150914 independently of other coincident noise investigations. 
Any PEM channel exhibiting power in the frequency band of GW150914 in excess of the expected maximum of Gaussian noise in a 1000-second interval was further examined. 
Two magnetometers at the Livingston detector sensitive to potential global coincident fields exhibited excess power at least 40 times too small to produce an event with the amplitude of GW150914.  No excess power was observed in any radio receivers. 

Given the global rate of lightning strikes, some coincidence with GW150914 is expected. 
The VAISALA GLD360 Global Lightning Dataset reported approximately 60 strikes globally during the second containing GW150914 \cite{Said;2013cur, Said;2010net}. 
One very strong lightning strike, with a peak current of about 500 kA, occurred over Burkina Faso (roughly 9,200~km from Livingston and 11,000~km from Hanford).
Fluxgate magnetometers indicate that magnetic disturbances at the LIGO detectors produced by coincident lightning strikes were at least 3 orders of magnitude too small to account for the amplitude of GW150914.

The PEM sensor network would easily detect any electromagnetic signal that would induce a transient in $h(t)$ with the same amplitude as GW150914. However, for redundancy, external observatories were also checked for natural or human-generated electromagnetic signals \cite{VLF, Moore, Learmonth, Callisto, RSTO, spaceweather, WIND, GOES, VIC} that coincided with GW150914. Geomagnetic signals at the time of the strike were estimated to produce $h(t)$ noise roughly 8 orders of magnitude smaller than the GW150914 signal at 100~Hz.

Although cosmic ray events are not expected to produce coincidences between detectors, the cosmic ray detector at LIGO-Hanford detected no events coincident with GW150914. Additionally, cosmic ray rates at the LIGO-Hanford site and external detectors around the world \cite{Ash, Bartole}  were low and exhibited no unusual fluctuations at the time of the event.


\subsection{Checks of auxiliary channels for noise coincident with GW150914}

Three algorithms are used to statistically identify correlations between transient noise identified in auxiliary channels and $h(t)$ for each detector \cite{Smith:2011, Isogai:2010, Essick:2013ov, Biswas:2013ap}.
Implementation details differ for each algorithm, but all work by defining a measure of correlation and identifying auxiliary channels with significant correlation relative to chance.

All three algorithms were effective in identifying correlations between transients in $h(t)$ and auxiliary channels by systematically removing a larger fraction of noise transients than the fraction of time removed for the week surrounding GW150914. 
Over the week surrounding GW150914, these algorithms successfully removed an average of ~6\% of noise transients at LIGO-Hanford and ~2\% at LIGO-Livingston for a deadtime of 0.1\%, which is 20-60 times greater than expected for chance coincidences.
None of the algorithms found a noise correlation within 180 seconds of the time of the event for LIGO-Livingston or within 11 seconds of the event for LIGO-Hanford. 

A comprehensive survey of transient excess power in all auxiliary channels was also conducted for at least 8 seconds around GW150914. 
Although no channel was statistically significant, a few of the transients nearby in time were followed up by hand in greater detail, as discussed in Section \ref{sec:ChanVet}. None were found to contribute to $h(t)$ in a way that might imitate or impact GW150914.

As part of a related check, auxiliary channels monitoring the control signals for optic motion actuation at both detectors were found to be well within their stable operating range at the time of GW150914. Consequently, even if an environmental perturbation were present it would not induce a transient in $h(t)$ due to control loop instability. 

\subsection{Vetting of channels with identified excess power near the event time}\label{sec:ChanVet}

\begin{figure}
\centering
\includegraphics[width=\linewidth]{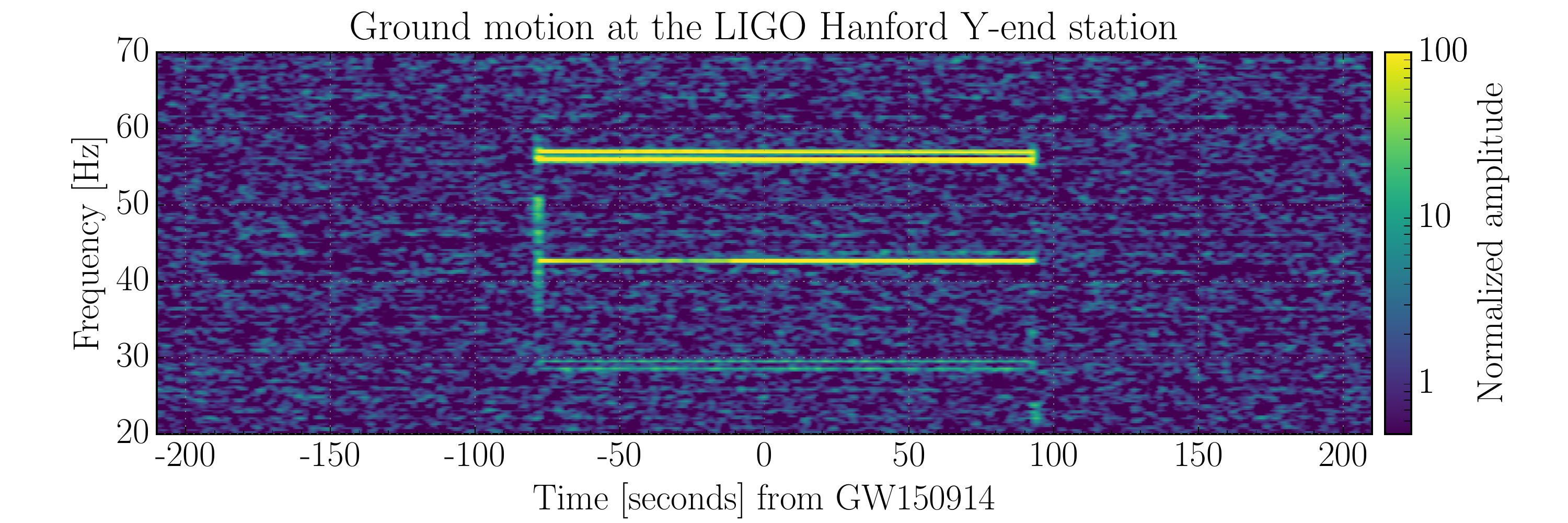}
\caption{A normalized spectrogram centered around the time of GW150914 of a Streckeisen STS-2 seismometer located near the Y-end test mass. An air compressor turns on at -75 seconds and off at +100 seconds. }
\label{fig:GW150914_aircompressor}
\end{figure}

A by-eye examination of spectrograms of every auxiliary channel identified a small subset of auxiliary channels that exhibited excess power within one second of GW150914, however, we found no evidence of noise that could generate GW150914 at either detector.
In addition to the magnetometer events discussed above in relation to potentially coincident sources, there were 4 excess power events identified in magnetometers that monitor electromagnetically noisy electronics rooms. The observed magnetic fields would have had to have been at least 20 times stronger  to account for the amplitude of GW150914 through coupling to the electronics. Channels from a seismometer and an accelerometer at LIGO-Hanford and two accelerometers at LIGO-Livingston also exhibited excess power.  These vibrational disturbances were at least 17 times too small to account for the amplitude of GW150914. None of the environmental events matched GW150914 in time and frequency behavior.  

The excess power triggers in the seismometer channels at LIGO-Hanford were likely due to a nearby air compressor with degraded vibration isolation that was running about 100m away from optical components during the detection of GW150914. This excess ground motion, shown in Figure  \ref{fig:GW150914_aircompressor}, lasted for approximately three minutes at multiples of about 14 Hz (28, 42, 56 Hz). During the second containing GW150914, the largest disturbance detected by the seismometer (at $\sim$56 Hz) was at least 30 times too small to account for the amplitude of GW150914.

There was also excess noise in the Livingston input mode cleaner \cite{aLIGO} that was ruled out as a potential indication of noise that might mimic GW150914. This noise had time-frequency morphology that was inconsistent with any potential coupling mechanism. In particular, all power was below 8~Hz and the noise duration was nearly one second. Such a long transient would be unlikely to couple from the input mode cleaner to $h(t)$ with duration comparable to GW150914 ($\sim$ 200 ms).

\subsection{Investigation of noise transients with similar morphology to CBC waveforms}\label{sec:noise chirps}

Both detectors occasionally record short noise transients of unknown origin consisting of a few cycles around 100~Hz, including blip noise transients, discussed in Section \ref{sec:noisesources}. 
None have ever been observed to occur in coincidence between detectors and follow-up examination of many of these transients confirmed an instrumental origin.  
While these transients are in the same frequency band as the candidate event, they have a characteristic time-symmetric waveform with significantly less frequency evolution, and are thus clearly distinct from the candidate event.  

\begin{figure}
  \centering
  \includegraphics[width=\textwidth]{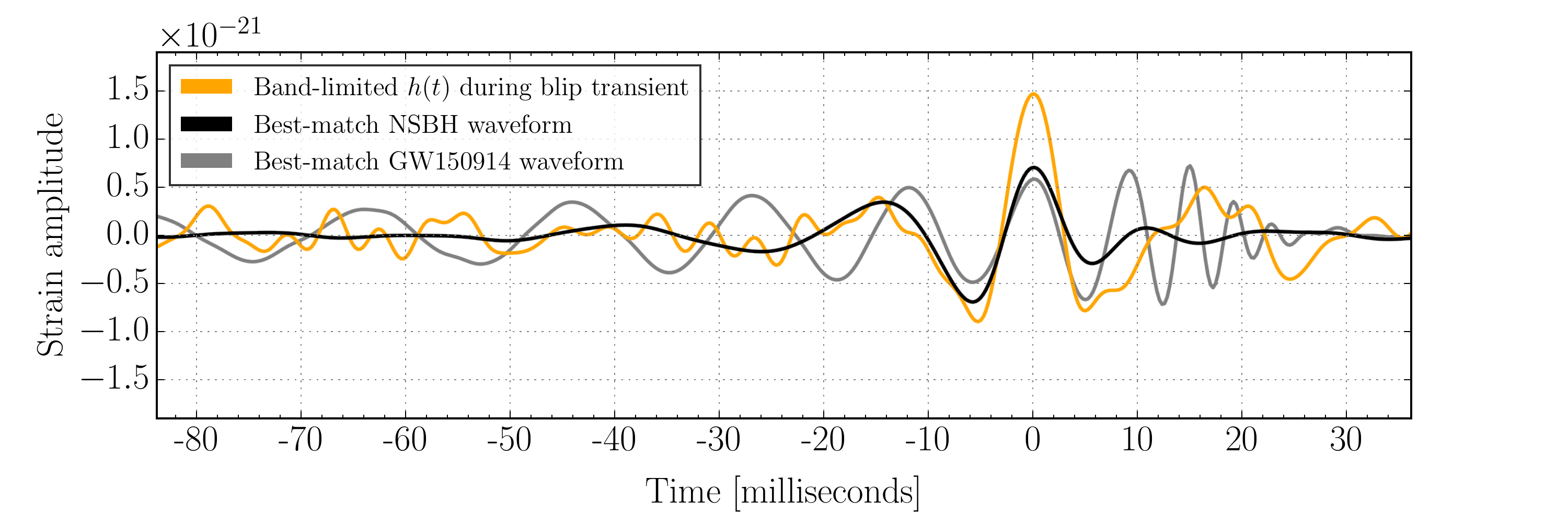}
  \caption{A blip transient in LIGO-Livingston strain data that produced a significant %
           background trigger in the CBC analysis in orange, %
           and the best-match template waveform (amplitude-scaled for %
           comparison) in black, which exhibits a few more low-SNR cycles but otherwise quite similar morphology. The best-match waveform for the GW150914 signal, in gray, is quite distinct from both the blip transient and the neutron-star-black-hole (NSBH) waveform that most closely matches it, with more than 10 distinct cycles shown and a significant increase in frequency over time. All three time series have the same zero-phase band-pass filter applied. 
           }
  \label{fig:blip-recon}
\end{figure}

To illustrate this, Figure \ref{fig:blip-recon} shows a blip transient that produced one of the most significant CBC background triggers associated with blip transients ($\chi^2$-weighted SNR $\gtrsim$ 9; compare to Figure \ref{fig:cbc-newsnr-histograms}) during the analysis period and the neutron-star-black-hole (NSBH) binary template waveform it most closely matched.  Although these noise transients do have significant overlap with regions of the CBC parameter space that produce very short waveforms, such as very high total mass binaries with extreme anti-aligned spins, they do not have a time domain morphology that matches CBC templates with similar character to GW150914.

The potential impact of any accidental coincidence between such noise transients on the sensitivity of the searches is accounted for in the reported background distribution. No noise transients identified to have similar morphology elements to CBC signals \cite{Powell:2015ona}, including blip transients, 
produced nearly as high a $\chi^2$-weighted SNR as GW150914. 


\subsection{LVT151012}\label{sec:G197392}
GW150914 was by far the most significant event in all transient search results over the sixteen days of analyzed data. The CBC search also identified the second most interesting event on the 12th of October 2015. 
This trigger most closely matched the waveform of a binary black hole system with masses \MONESCOMPACTSecondMonday\ M$_{\odot}$ and \MTWOSCOMPACTSecondMonday\ M$_{\odot}$, producing a trigger with a false-alarm rate of 1 event per \CBCSECONDEVENTIFAR\ years; far too high to be a strong detection candidate
\cite{Aasi:2016bh, Aasi:2016cbc, Aasi:2016pe}. 

\begin{figure}[!ht]%
\centering
  \includegraphics[width=\textwidth]{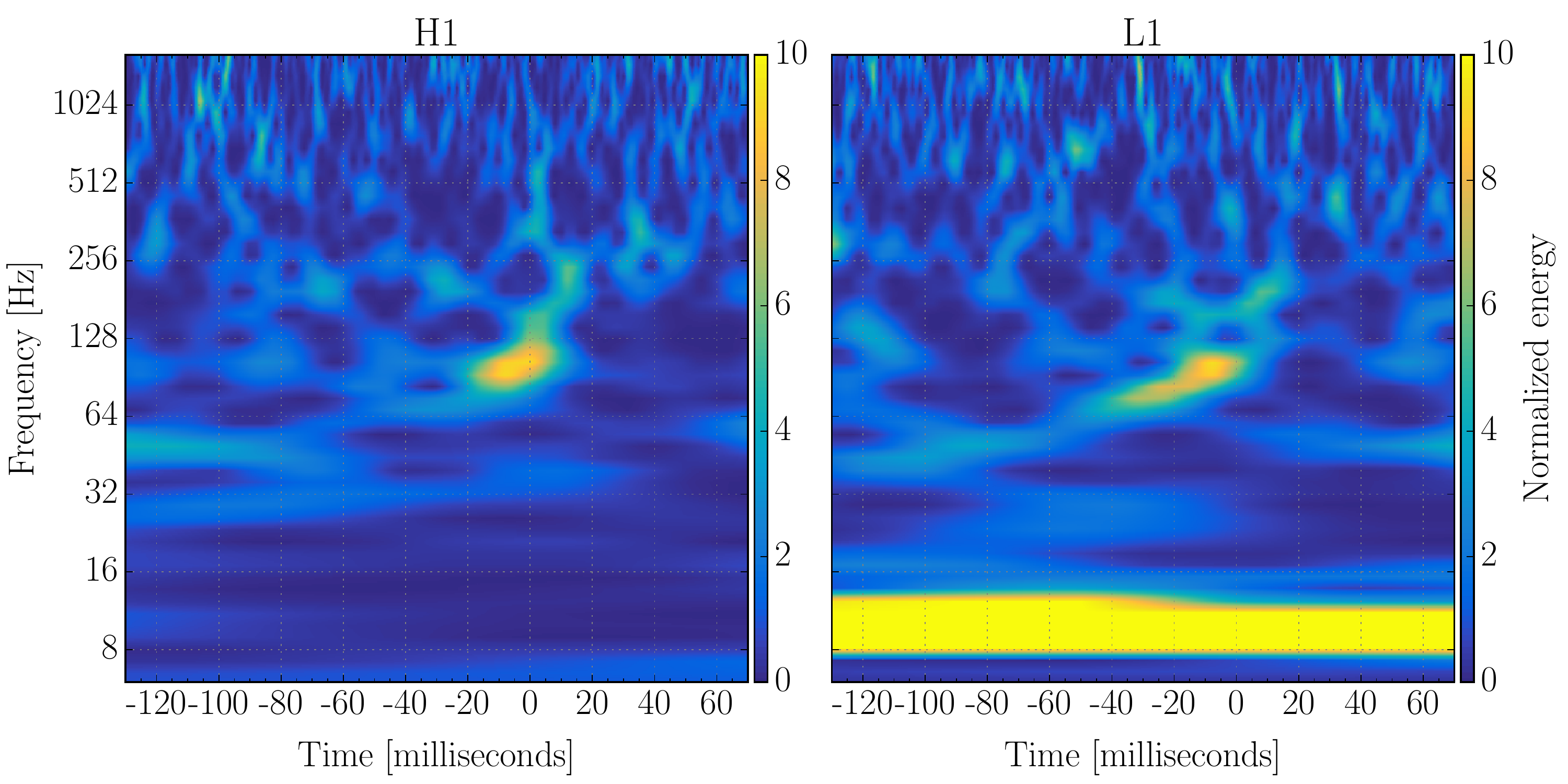}
  \caption{Normalized spectrograms of LVT151012 in LIGO-Hanford (left) and LIGO-Livingston (right) $h(t)$ data with the same central GPS time. Note these spectrograms have a much smaller normalized energy scale than those in Figure \ref{fig:GW150914_omegascans}. }
\label{fig:G197392_omegascans} 
\end{figure}

We performed similar in-depth checks of potential noise sources for this trigger. For LIGO-Livingston data, LVT151012 is in coincidence with significant excess power at 10Hz lasting roughly three seconds, a portion of which can be seen in Figure \ref{fig:G197392_omegascans}.
 There is no obvious indication of upconversion to the frequency range analyzed by the transient searches, so the low frequency noise is not thought to have caused the signal associated with LVT151012 in the Livingston detector. 
 
 The data around this event were found to be significantly more non-stationary than those around 
GW150914. The noise transient rate in the hours around LVT151012 was significantly higher than usual at both LIGO detectors, seen in the Omicron trigger rate even on a broad time scale for LIGO-Livingston in particular, as illustrated in Figure \ref{fig:aligo-glitch-rate}. 
This was likely due to increased low frequency ground motion associated with ocean waves \cite{Daw:2004sei}.
The elevated noise transient rate at both sites induced a higher rate of background triggers around the time of LVT151012. 

No detector characterization studies to date indicate that LVT151012 was caused by a noise artifact. 

\subsection{Noise transient rate}

\begin{figure}
  \centering
  \includegraphics[width=\linewidth]{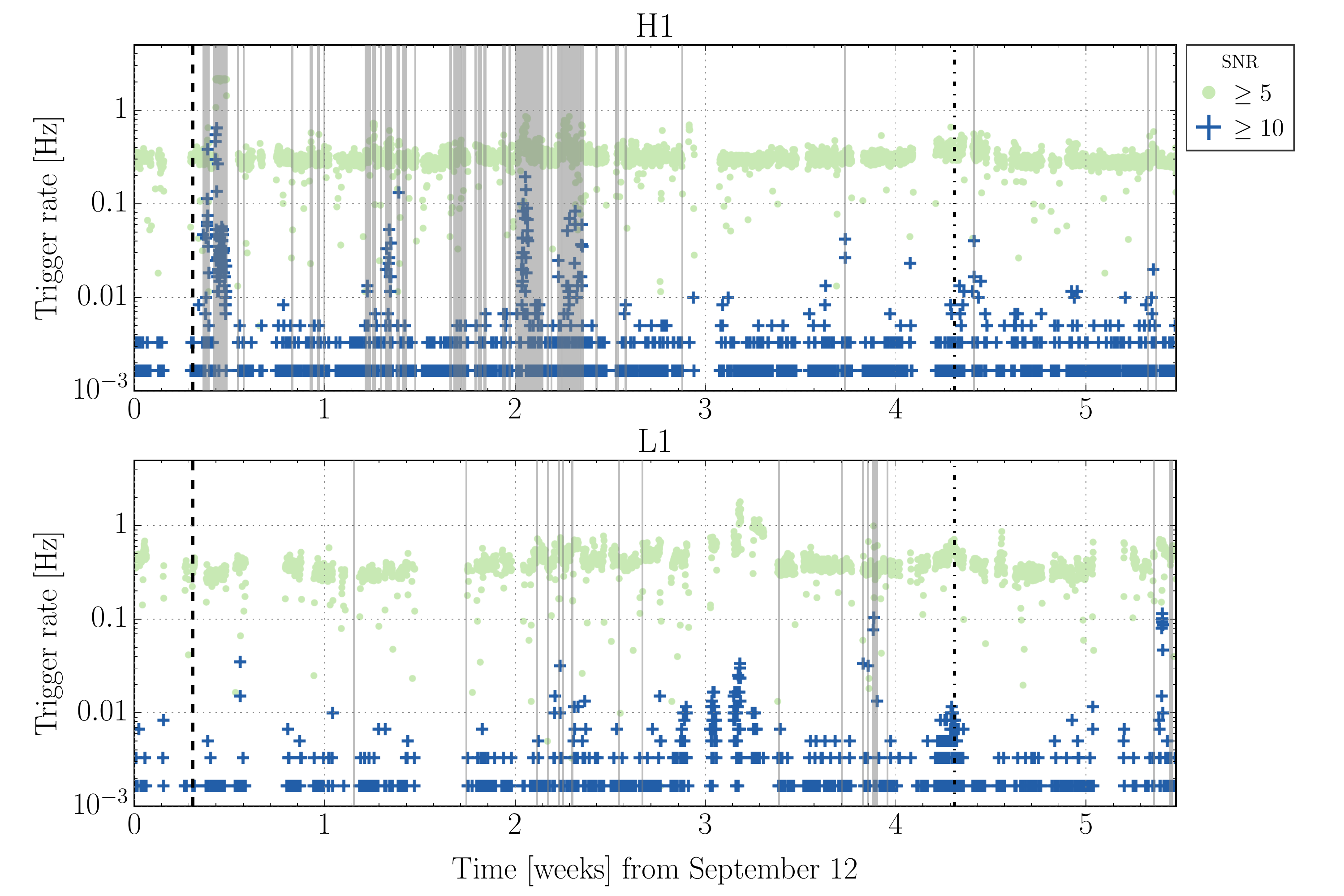}
  \caption{The rate of transient noise as witnessed by the single detector burst algorithm Omicron for the LIGO Hanford (above) and LIGO-Livingston (below) detectors. Each dot represents the average trigger rate over a 600 second interval. Green dots show triggers with an SNR above 5, and blue crosses show triggers with an SNR above 10. Time vetoed from the analysis period is indicated in gray. The time of GW150914 is indicated with a vertical dashed line and LVT151012 with a dot-dashed line.}
  \label{fig:aligo-glitch-rate}
\end{figure}

\Cref{fig:aligo-glitch-rate} shows the rate of transient noise in the data as identified by the single-detector burst algorithm Omicron for each of the two detectors over the analyzed period. GW150914 occurs during a period when the transient noise rate is low at both detectors, particularly for louder transient noise. However, event LVT151012 occurs during a period when the rate of transient noise is elevated, likely due to increased seismic noise, as described below.  

For LIGO-Hanford, major excursions from the normal noise transient rate of $\sim0.3$\,Hz can be seen around 3 days into the analysis period due to an electronics failure in the instrumental control system; similarly smaller problems are seen in the second and third weeks due to problems with high seismic noise, and faulty radio frequency modulation electronics as described in \ref{sec:RF45}. 
Periods with a significantly elevated noise transient rate at the Hanford detector are largely removed from the analyzed period by the category 1 data quality veto associated with these faulty electronics. 
For LIGO-Livingston, a high noise transient rate is observed throughout weeks three and four, due in part to poor weather conditions and elevated seismic noise. The instrumental coupling was not well enough understood to generate an effective data quality veto for this elevated noise.

\section{Conclusions}\label{sec:concl}

At the time of GW150914, the LIGO detectors were operating in a low-noise state with nominal environmental and instrumental noise. 
Following the event, the detectors were maintained in the same configuration to ensure that detector changes would not cause unanticipated consequences which might bias the background estimation for the event. 
The backgrounds measured by the transient searches were stable throughout this analyzed period. 
Data quality vetoes were produced for each detector in response to instrumental or environmental noise sources.
We conclude that the selected analysis period provides an accurate estimation of the significance of GW150914. 

Additionally, thorough investigations found no evidence that environmental influences or non-Gaussian detector  noise  at either LIGO site might have caused the observed gravitational wave signal GW150914.
A detailed study of environmental influences conclusively ruled out all postulated potential sources of correlated detector output at the time of the event, except for a binary black hole gravitational wave signal. 

Characterization of the LIGO detectors via investigations of noise types that most impact the astrophysical searches and mitigation of noise couplings will continue to play a critical role in gravitational wave astronomy. Reducing the rate of high-significance background events and increasing search sensitivity is particularly important for near-threshold events such as LVT151012. Detector characterization will effectively expand the range of astrophysical sources that the gravitational wave detectors are sensitive to, providing a significantly greater number, and perhaps also variety, of events from which we can draw confident physical inferences.  

%
\section{Acknowledgements}
The authors gratefully acknowledge the support of the United States
National Science Foundation (NSF) for the construction and operation of the
LIGO Laboratory and Advanced LIGO as well as the Science and Technology Facilities Council (STFC) of the
United Kingdom, the Max-Planck-Society (MPS), and the State of
Niedersachsen/Germany for support of the construction of Advanced LIGO 
and construction and operation of the GEO600 detector. 
Additional support for Advanced LIGO was provided by the Australian Research Council.
The authors gratefully acknowledge the Italian Istituto Nazionale di Fisica Nucleare (INFN),  
the French Centre National de la Recherche Scientifique (CNRS) and
the Foundation for Fundamental Research on Matter supported by the Netherlands Organisation for Scientific Research, 
for the construction and operation of the Virgo detector
and the creation and support  of the EGO consortium. 
The authors also gratefully acknowledge research support from these agencies as well as by 
the Council of Scientific and Industrial Research of India, 
Department of Science and Technology, India,
Science \& Engineering Research Board (SERB), India,
Ministry of Human Resource Development, India,
the Spanish Ministerio de Econom\'ia y Competitividad,
the Conselleria d'Economia i Competitivitat and Conselleria d'Educaci\'o, Cultura i Universitats of the Govern de les Illes Balears,
the National Science Centre of Poland,
the European Commission,
the Royal Society, 
the Scottish Funding Council, 
the Scottish Universities Physics Alliance, 
the Hungarian Scientific Research Fund (OTKA),
the Lyon Institute of Origins (LIO),
the National Research Foundation of Korea,
Industry Canada and the Province of Ontario through the Ministry of Economic Development and Innovation, 
the Natural Science and Engineering Research Council Canada,
Canadian Institute for Advanced Research,
the Brazilian Ministry of Science, Technology, and Innovation,
Funda\c{c}\~ao de Amparo \`a Pesquisa do Estado de S\~ao Paulo (FAPESP),
Russian Foundation for Basic Research,
the Leverhulme Trust, 
the Research Corporation, 
Ministry of Science and Technology (MOST), Taiwan
and
the Kavli Foundation.
The authors gratefully acknowledge the support of the NSF, STFC, MPS, INFN, CNRS and the
State of Niedersachsen/Germany for provision of computational resources.

\section{References}
\bibliographystyle{unsrt}
\bibliography{references}

%

\appendix 

\section{Example data quality veto: 45 MHz light modulation transients}\label{sec:RF45}

\begin{figure} [!ht]
\centering
\includegraphics[width=\textwidth]{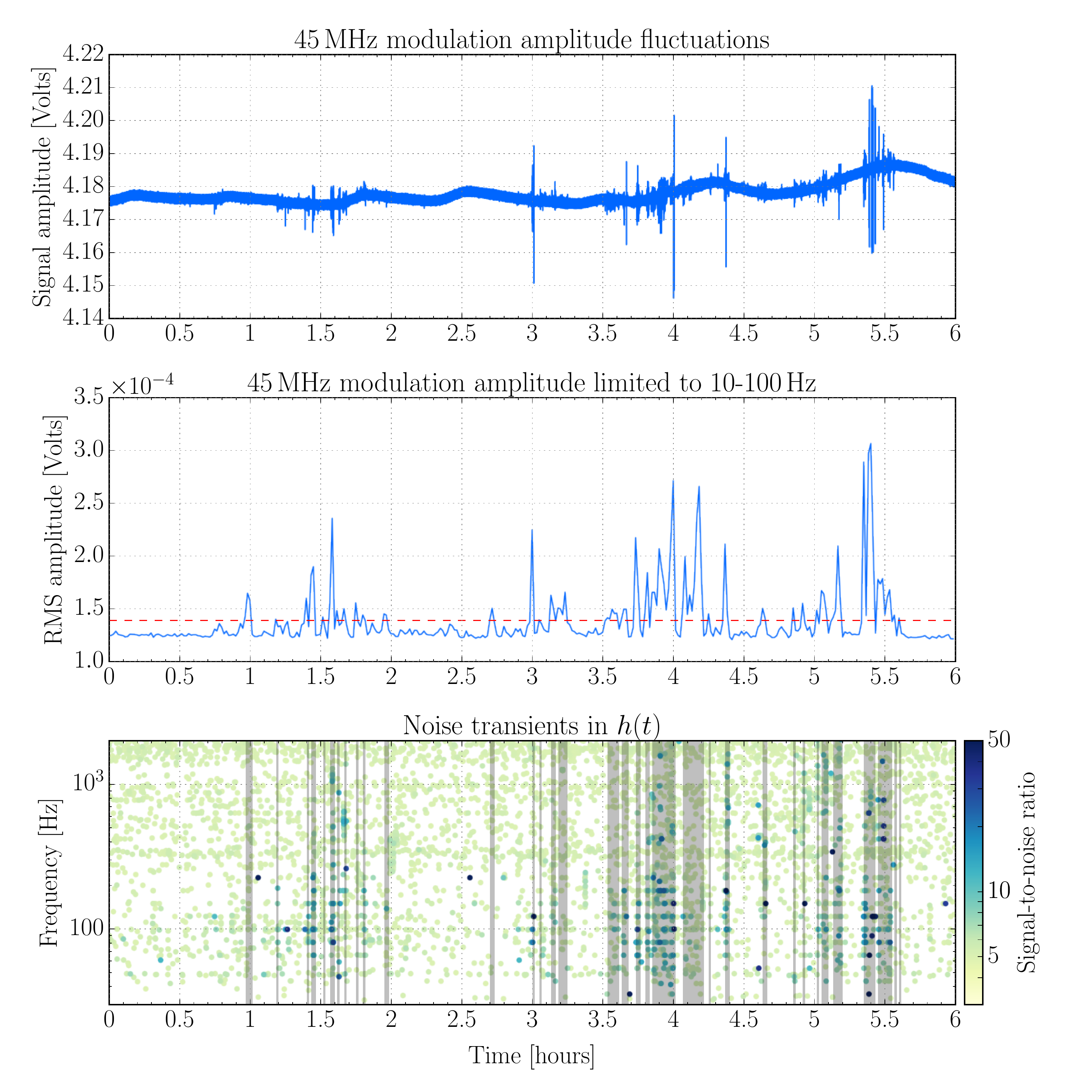}
\caption{The effectiveness of the veto criteria designed to flag $h(t)$ non-stationarity due to the malfunction of the 45MHz driver over a six hour period on September 21, 2015. 
The top panel shows the witness channel (a monitor for amplitude fluctuations in the signal used to generate the 45 MHz optical sidebands) over a 6 hour period with non-stationary data in $h(t)$. 
Due to variation in its mean value, a band limited root-mean-square (BLRMS) of this channel over 60 seconds was a better indicator of the targeted behavior, shown in the middle panel. 
Thresholds of this BLRMS were tested over 11 days during the analysis period for efficiency in identifying periods of high trigger rate in $h(t)$, and the threshold shown in the middle figure was found to be optimal for the analysis time removed. 
The bottom panel shows Omicron $h(t)$ triggers over the same 6 hour time period. Times removed by the veto are shaded out in gray.}
\label{fig:RF45ts}
\end{figure}

A data quality veto is generally constructed using an auxiliary channel which is strongly correlated with an instrumental problem. 
A notable example from the analyzed period was observed at LIGO-Hanford; intermittent periods with a significantly elevated transient noise rate in $h(t)$. 
This behavior began suddenly five days before GW150914, independent of any activities taking place on site. 
The behavior was traced back to the 45 MHz electro-optic modulator driver system used to generate optical cavity control feedback signals \cite{aLIGO}. 
To find the auxiliary channel which best correlates with non-stationary data in $h(t)$, auxiliary channels recording interferometric cavity readouts and control signals associated with this driver were examined for excursions coincident with $h(t)$ noise transients. 
A channel monitoring amplitude fluctuations in the signal used to generate the 45 MHz optical sidebands was found to be the best indicator of this non-stationary behavior. 

\begin{figure} [!ht]
\centering
{\includegraphics[width=.5\textwidth]{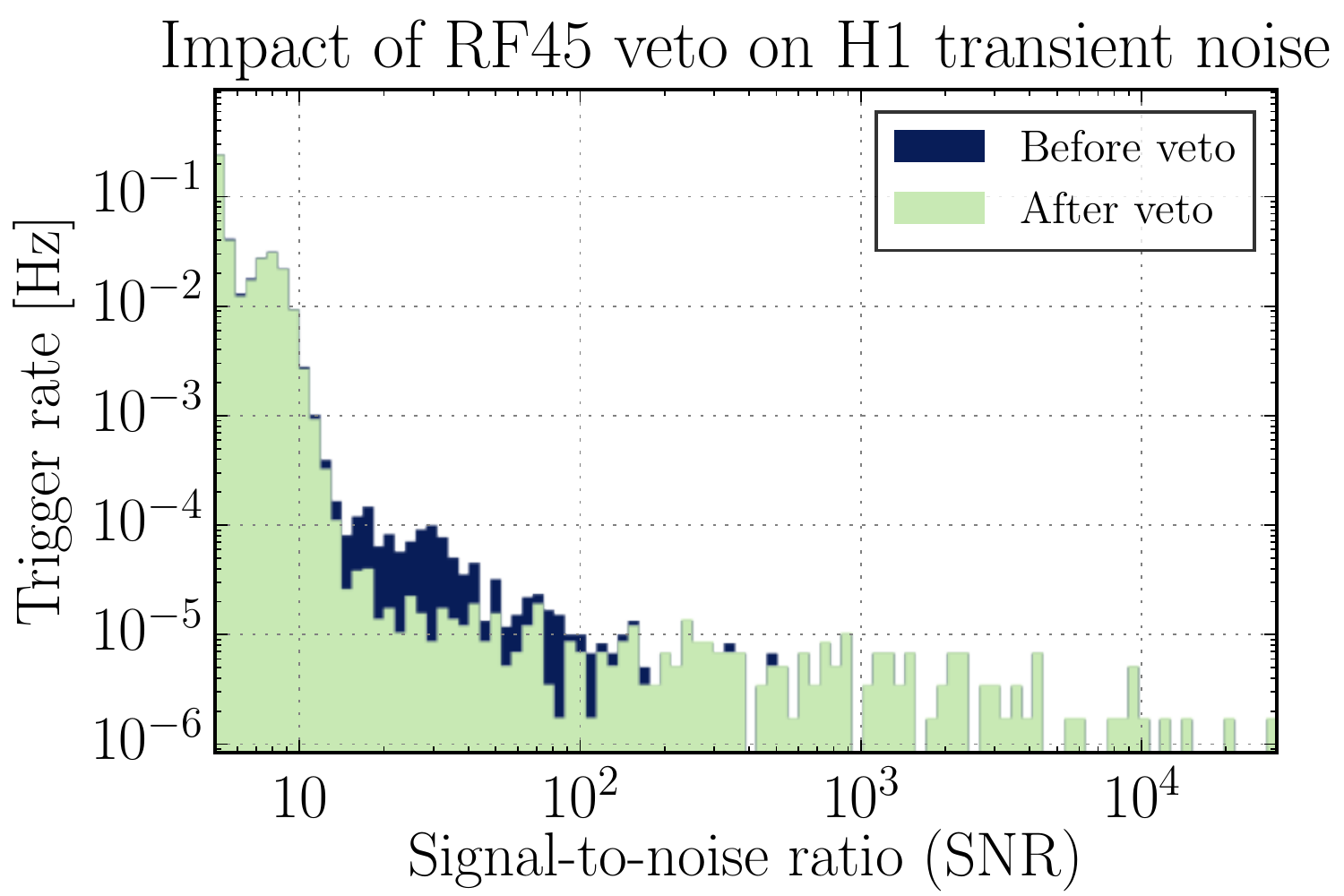}}
\caption{\label{RF45veto}The rate of Omicron triggers with and without
vetoes applied to 11 days of data, a subset of the analysis period. The veto 
is effective at removing excess triggers with a SNR between 15 and 100. When applied to the full GW150914 analysis period, this data quality veto removed 42$\%$ of noise transients of an SNR of 20 or greater, at the expense of 2.6$\%$ of coincident data.}
\label{fig:RF45trig_rate}
\end{figure}

Spikes in this auxiliary channel correlate well with a high rate of noise transients seen in $h(t)$. However, the mean value of this channel varies significantly over time, meaning a simple threshold on the timeseries was not suitable for defining a data quality veto. 
Instead, band-limited root-mean-square values of this witness channel over minute strides were used. The effectiveness of different thresholds was tested using an 11 day subset of the analysis period. 
An example of the behavior of this veto over a 6 hour time scale can be seen in Figure \ref{fig:RF45ts}.  With the selected threshold, this data quality veto removed 56$\%$ of noise transients with a SNR $>$ 20, while only introducing 3$\%$ of deadtime over the 11 days of data. 
Figure \ref{fig:RF45trig_rate} shows the distribution of Omicron triggers identified and removed, over 
the 11 days, by this veto. 

This data quality flag was applied as a category 1 veto to the transient gravitational wave searches, responsible for removing 2.62$\%$ of the total coincident time from the analysis period. 

\section{The physical environment monitor (PEM) array}\label{sec:PEM}

The environment can influence the detector by mechanical force, electromagnetic waves, static electric and magnetic fields, and possibly high-energy radiation from cosmic rays.  
Mechanical forces, due to ground motion, temperature fluctuations, or air pressure fluctuations, are transmitted through structures that house and support interferometer optics and other key instrumentation.

Certain global-scale environmental effects could influence both detectors within 10 ms, which is the light travel time between the LIGO detectors and the maximum time delay for a gravitational wave signal of astrophysical origin.   
A network of sensors is employed such that global-scale environmental disturbances that could influence the detectors, such as electromagnetic disturbances in the atmosphere or transient fluctuations in the power grid, are redundantly monitored using PEM sensors that are significantly more sensitive to these disturbances than the detectors themselves.  

By monitoring the immediate environment for disturbances that can be transmitted to the detector strain signal, we cover a large variety of environmental effects that can influence the detector data. 
For example, wind can couple through vibrations in the ground and air, and its behavior is witnessed by seismometers, accelerometers, and microphones (audio and infrasound frequencies). 
Lightning could couple by magnetic fields and electromagnetic waves at frequencies that we demodulate into the detection band for optic cavity control \cite{aLIGO} and is monitored by magnetometers and radio frequency receivers.

\begin{figure}
\includegraphics[width=\linewidth]{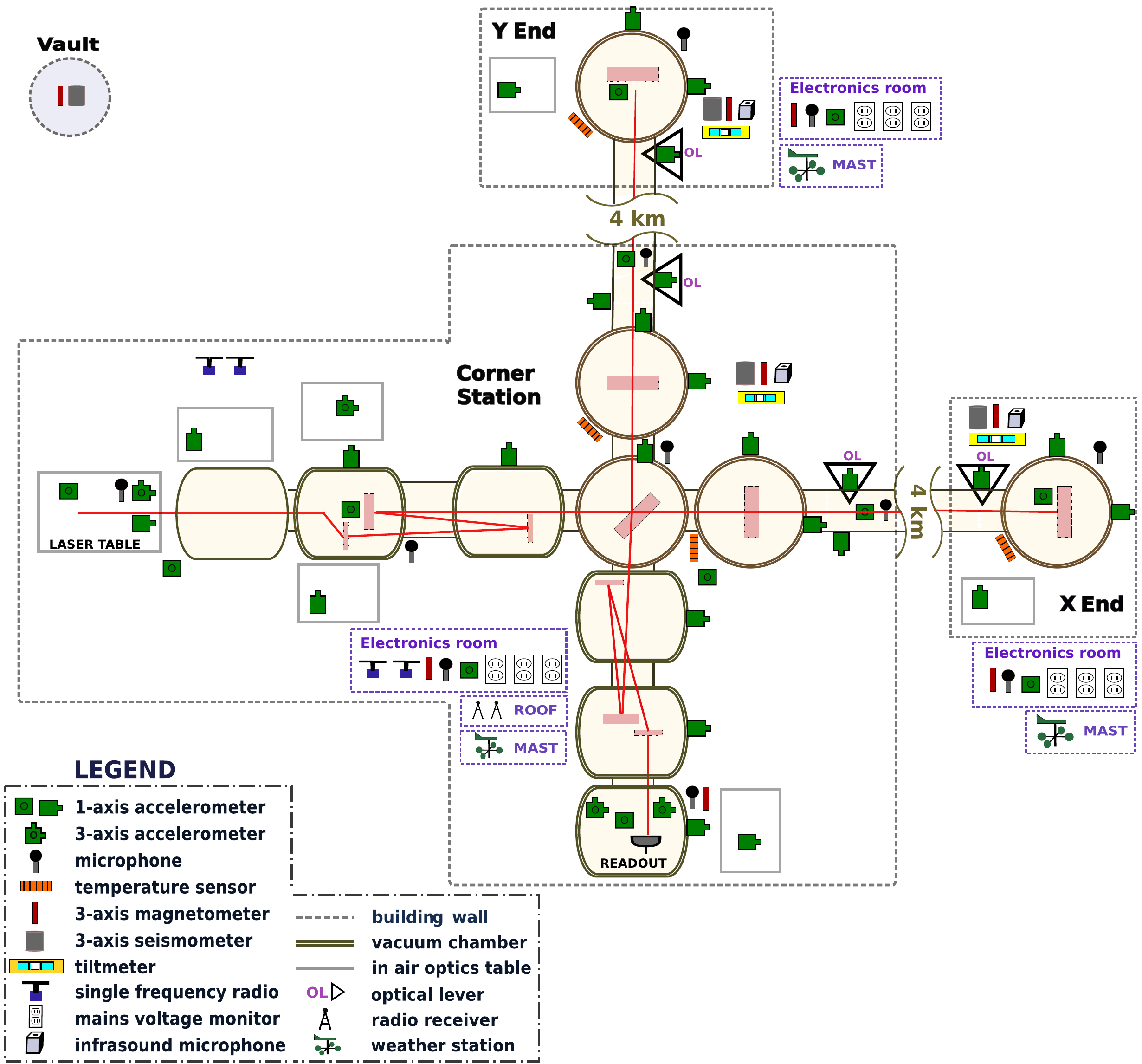}
\caption{The physical environment monitor (PEM) array at the Livingston detector, as seen on http://pem.ligo.org \cite{PEM}. 
Gray dashed lines enclose instrumentation in separate structures: the corner station building located at the vertex of the laser-interferometric detector, the two end stations located at the end of the 4km detector arms, and the `vault', which houses PEM sensors away from all buildings to measure noise due to the external environment. 
Purple dashed lines indicate rooms within structures, or spaces just outside of structures. For example, the corner station and both end stations have PEM sensors in electronics rooms containing computers that sense and control the detector as well as PEM equipment mounted on a mast on the roof. 
See \cite{Aasi:2016in, aLIGO} for detailed description of the optical layout shown. 
}
\label{fig:PEM}
\end{figure}

Figure \ref{fig:PEM} shows how these sensors are distributed at key locations throughout the LIGO-Livingston detector site (the LIGO-Hanford layout is very similar) \cite{PEM}. 
Each building is equipped with seismometers and ground tilt sensors to monitor the motion of the concrete slab on which vacuum chambers and optical tables are mounted. 
Each of these buildings also contains an infrasound microphone and a set of audio-frequency microphones, including a microphone near the electronics that control the detector feedback loops and acquire auxiliary channel data. 
Power voltage monitors are installed in the electronics room of each building. 
Fluxgate magnetometers sense disturbances in the local magnetic field in all electronics rooms as well as a nearby subset of vacuum chambers. 
Accelerometers are mounted on vacuum chamber walls as well as on in-air optics tables and the concrete slab of each building. 
External to the detector buildings are radio frequency receivers as well as wind speed sensors and outdoor weather stations. 
The PEM system at the Hanford detector includes a cosmic ray detector located underneath one of the test masses.

There are a total of 173 PEM channels at LIGO-Hanford and 130 at LIGO-Livingston, where a greater number of channels at Hanford is due to additional redundancy in sensors as well as the cosmic ray detector. 

\end{document}